\documentclass[%
 reprint,
 amsmath,amssymb,
 aps,
]{revtex4-1}

\usepackage[english]{babel}
\usepackage{graphicx}% Include figure files
\usepackage{dcolumn}% Align table columns on decimal point
\usepackage{bm}% bold math
\usepackage{wrapfig}
\usepackage{comment}
\usepackage{xcolor}
\definecolor{DarkGreen}{rgb}{0.0, 0.5, 0.0}
\newcommand{\up}{\uparrow}
\newcommand{\down}{\downarrow}
\newcommand{\G}{\mathcal{G}}
\usepackage{hyperref}

\begin{document}

\title{Exact dynamics of quantum systems driven by time-varying Hamiltonians: solution for the Bloch-Siegert Hamiltonian and applications to NMR}

\author{Pierre-Louis Giscard}%
 \email{giscard@univ-littoral.fr}
\affiliation{%
Universit\'e du Littoral C\^ote d'Opale, EA 2597, Laboratoire de Math\'ematiques Pures et Appliqu\'ees Joseph Liouville, F-62228 Calais, France
}%

\author{Christian Bonhomme}%
 \email{christian.bonhomme@upmc.fr}
\affiliation{%
Laboratoire de Chimie de la Mati\`ere Condens\'ee de Paris, Sorbonne Universit\'e, UMR CNRS 7574, 4, place Jussieu, 75252, Paris Cedex 05, France
}%

\date{\today}
             
\begin{abstract}
Comprehending the dynamical behaviour of quantum systems driven by time-varying Hamiltonians is particularly difficult. Systems with as little as two energy levels are not yet fully understood as the usual methods including diagonalisation of the Hamiltonian do not work in this setting. In fact, since the inception of Magnus' expansion in 1954, no fundamentally novel mathematical approach capable of solving the quantum equations of motion with a time-varying Hamiltonian has been devised. We report here of an entirely different non-perturbative approach, termed path-sum, which is always guaranteed to converge, yields the exact analytical solution in a finite number of steps for finite systems and is invariant under scale transformations of the quantum state space. Path-sum can be combined with any state-space reduction technique and can exactly reconstruct the dynamics of a many-body quantum system from the separate, isolated, evolutions of any chosen collection of its sub-systems. As examples of application, we solve analytically for the dynamics of all two-level systems as well as of a many-body Hamiltonian with a particular emphasis on NMR (Nuclear Magnetic Resonance) applications: Bloch-Siegert effect, coherent destruction of tunneling and $N$-spin systems involving the dipolar Hamiltonian and spin diffusion. 
\end{abstract}


\maketitle

\section{Introduction}
The unitary evolution operator $\mathsf{U}(t',t)$ describing the time dynamics of quantum systems is defined as the unique solution of Schr\"odinger's equation with quantum Hamiltonian $\mathsf{H}$, i.e. $(\hbar=1)$
\begin{equation}\label{DefU}
-i\,\mathsf{H}(t')\,\mathsf{U}(t',t)=\frac{d}{dt'}\mathsf{U}(t',t),
\end{equation}
and such that $\mathsf{U}(t'=t,\,t)=\mathsf{Id}$ is the identity matrix at all times. Evidently, this operator plays a crucial role at the heart of quantum mechanics, including for spin dynamics in Nuclear Magnetic Resonance (NMR) \cite{slichter1990,ernst1987,Mehring2001}. 
%When the quantum Hamiltonian is time-independent, the evolution operator takes on the mathematically simple and compact form  $\mathsf{U}(t',t)=\exp[-i(t'-t)\mathsf{H}]$. In the general case however, and 
As is typically the case in NMR, the Hamiltonian may be time-dependent and might furthermore not commute with itself at various times, $\mathsf{H}(t)\mathsf{H}(t')-\mathsf{H}(t')\mathsf{H}(t)\neq \mathbf{0}$ for $t'\neq t$. In this situation, the evolution operator no longer has a simple calculable form in terms of the Hamiltonian, e.g. \emph{it cannot even be evaluated via direct diagonalisation} of $\mathsf{H}$. Rather, $\mathsf{U}$ is formally described by the action of a time-ordering operator on the Dyson series representation of the quantum evolution \cite{dyson1952}, a formulation which does not permit concrete calculations to be carried out.
%the action of a time-ordering operator on the exponential of the Hamiltonian \cite{dyson1952}
%, an expression first expounded by Dyson \cite{dyson1952} but which is little more than a notation precluding immediate evaluations. 

%Whereas this problem has been discussed in details in the NMR literature, calculations of the evolution operator remain barely tractable for an in-depth description of spin dynamics. 
As a consequence, only approximate expressions of $\mathsf{U}(t',t)$ are obtained  and these are only accurate for short times. 
%The same conclusion holds for the so called product representation of the Hamiltonian involving its separation in two parts of equal importance \cite{Mehring2001}.  
A major breakthrough in the description and understanding of solid state NMR was %achieved 
%by 
the inception of Average Hamiltonian Theory \cite{Waugh1968}. %\cite{Waugh1968, Haeberlen1976}. 
This relies exclusively on the Magnus expansion \cite{Magnus1954} of $\mathsf{U}(t',t)$. 
%Discovered in 1954, this expansion is the last essentially new development 
%in the mathematical theory of time-dependent Hamiltonians. 
%ME was the starting point for a new era in homonuclear decoupling \cite{Haeberlen1968,Burum1979} experiments and 
%multiple-quantum spectroscopy \cite{Warren1980}. 
%The fundamental idea behind the Magnus expansion relies on  calculating a matrix whose `true' exponential gives the evolution operator, a formulation which involves an infinite series of time-dependent nested commutators of the Hamiltonian, mathematically a continuous form of the celebrated Baker-Campbell-Haussdorff formula \cite{bialynicki1969}. %Campbell1897, Baker1905, Hausdorff1906,
%Such terms can be represented in a compact form by the Bialynicki-Birula formula \cite{bialynicki1969}. 
However, higher order terms of the series remain highly cumbersome to write down explicitly so that practically, only low orders of the expansion are useable. Most importantly, Magnus expansion suffers from severe and incurable 
divergences as already mentioned by Magnus \cite{Magnus1954} and Fel'dman  \cite{Feldman1984}. %in early contributions \cite{Maricq1982}. Quoting Mehring and Weberruss in \cite{Mehring2001}: ``\emph{The convergence of this 
%series is of concern and must be considered in special cases}'' (!). The specific problem of 
%convergence has been re-investigated in-depth recently \cite{Blanes2009,Moan2008} and saturated upper bounds for the largest times reachable by ME before divergence have been discovered which severely restrict its use to short times \cite{Sanchez2011}. 
In the more specific case of periodically time-dependent Hamiltonian, such as those encountered in Magic Angle 
Spinning (MAS) experiments, it is well known that Magnus expansion suffers from a further two limitations, i.e. the stroboscopic detection of the NMR events, and the impossibility to take into account 
more than one characteristic period. 
%We mention that in the case of periodic Hamiltonian, 
%Secular Average Theory (SAT) can be applied as well \cite{Mehring2001}). %\cite{mehring1983, Goldman1993}. 
%It implies the separation of constants terms in the Hamiltonian avoiding non-secular contributions in first/higher order terms of the expansion. Using AHT, such terms may lead to erroneous results \cite{Goldman1992}. Ultimately, the 
%SAT approach follows the general concept developed in ME, with $\mathsf{U}(t',t)$ given by the `true' exponential of an infinite series of operators.

In the case of periodic Hamiltonian, Floquet theory dictates %(FT) holds 
%\cite{Floquet1883} %and implies 
that the evolution operator takes on the form $\mathsf{U}(t,0)=\mathsf{P}(t)\exp(\mathsf{F} t)$, with $\mathsf{P}(t)$ a periodic time-dependent matrix and $\mathsf{F}$ a constant matrix, both of which are determined perturbatively when working analytically \cite{Blanes2009}, or otherwise via numerical procedures \cite{Grossmann1992,grifoni1998}. Floquet formalism was first used by Shirley \cite{Shirley1965}, %whose fundamental contribution was to apply 
%Floquet theory 
who applied it to the case of a linearly polarised excitation in magnetic resonance and 
to give low orders analytical expressions for the Bloch-Siegert effect 
\cite{Bloch1940}. 

We also mention numerical methods: (i) Fer and Magnus-Floquet hybrids  proposed recently as potential expansions for the evolution operator \cite{Takegoshi2015,Mananga2018}, (ii) Zassenhaus and Suzuki-Trotter propagator approximations \cite{Brusch1997,Dumez2010,Mentink2017}. 
%in various contexts 
%(spin diffusion between dipolarly coupled like-spins and DNP).
The expansions presented above all suffer from various drawbacks including: the divergence of the series at long times; the perturbative nature of the numerical or theoretical approach; the non-avoidable propagation of errors at long time; the failure to find exact solutions even for small, one spin $1/2$, $2\times 2$ Hamiltonians. See also Appendix~\ref{AppA} for more litterature background.
%Finally, while our modest introduction gives an overview of the state-of-the-art, it is beyond the scope of the present article to cite all relevant works pertaining to quantum systems with time-varying Hamiltonians (see remarks in SI). Instead, we believe that a review article on the matter is required and would be beneficial to all involved. 

In this contribution, the path-sum method is applied for the very first time to NMR Hamiltonians to determine the corresponding evolution operators $\mathsf{U}(t',t)$. The rigorous underpinnings of this approach were laid out in \cite{Giscard2015,Giscard2012} within the general mathematical framework of systems of coupled linear differential equations with non-constant coefficients. So far, no  physical applications of these works has been presented. Consequently, they remained unnoticed outside of a specialised mathematical community, and their applicability to long-standing questions pertaining to quantum systems driven by time-dependent Hamiltonians went completely unrecognised.
It thus appears important to introduce path-sum to the physics community via illustrative examples bearing directly on currently open problems. 
%and the general lack of fundamental progress on the theory since Magnus' work in 1954;% \cite{Magnus1954}; 
%We focus on NMR in the following.
%We may put path-sum in a broader physical context by quoting R. P. Feynman \cite{Feynman2005}: ``\emph{With application to quantum mechanics, path integrals suffer most grievously from a serious defect. They do not permit a discussion of spin operators or other such operators in a simple and lucid way}''. From our point of view, path-sum precisely achieves what path-integrals could not for spin systems and go further by performing formal resummations on the infinitely many diagrams. Indeed, in the Hamiltonian formalism of spin systems, the quantum state space takes on the form of a discrete graph and its walks, weighted by the energy functional, are the analogs of the Feynman diagrams, all of which are re-summed via a single finite continued fraction over a few \emph{prime} walks. As a corollary, each of the finitely many term of the path-sum continued fraction represents a fundamental physical process from which all possible processes stem via nestings (insertion of a process into another).  While these interpretations are correct and appealing, path-sum's validity is independent from quantum theory and physics in general: it is rather a fundamental property of graph walks valid for even the most abstract systems of coupled linear differential equations with non-constant coefficients. 
Overall, it appears that the present work is the first fundamentally new approach to the problem of simulating quantum dynamics induced by time-varying Hamiltonians since Magnus' 1954 seminal results.

Path-sum is firmly established on three fundamentally novel concepts, insofar never applied within the quantum physics framework: (i) the representation of $\mathsf{U}(t',t)$ as the inverse of an operator with respect to certain $\ast$-product;
(ii) a mapping between this inverse, and sums of weighted walks on a graph; and (iii) fundamental results on the algebraic structure of sets of walks which exactly transform any infinite sum of weighted walks on any graph into a single branched continued fraction of finite depth and breadth with finitely many terms. Taken together, these three results imply that, for finite dimensional Hamiltonians, any \emph{entry} or \emph{block of entries} of $\mathsf{U}(t',t)$ has an exact, unconditionally convergent analytical expression that always involves a finite number of terms. We emphasise that throughout this work, the time $t$ is and remains a \emph{continuous} variable, in particular path-sum does not rely on time-discretisation. %involving in terms of $\ast$-products and inverses of time-dependent \emph{functions} (rather than matrices). 
%These inverses are themselves determined by unconditionally convergent analytical series or can be found via numerical tools pertaining to linear Volterra integral equations of the second kind.% with separable kernels. 
%Since every piece of the evolution operator is obtained exactly after a finite number of operations, the method is necessarily convergent. 
%Moreover, convergence is super-exponential in the number of terms retained in the PS continued fraction. 
As a corollary, path-sum yields a non-perturbative formulation of $\mathsf{U}(t',t)$, as will be illustrated below with the Bloch-Siegert effect. Further properties of path-sums ensures its scalability to multi-spin systems, for example allowing it to recover the exact dynamics of an entire system from the separate, isolated, evolutions of any chosen collection of its sub-systems. In its general form, path-sum is best understood as a method to exactly and analytically solve systems of coupled linear differential equations with non-constant coefficients.\\[-.7em]  

This article is structured as follows. We first present the mathematical background of \cite{Giscard2015} culminating in the path-sum formulation of quantum dynamics.  
%The fundamental scalability property of path-sum is introduced as well. 
In a second part, we detail applications in connection with general quantum theory and then more specifically with NMR. The first one provides the \textit{general} solution of Schr\"odinger's equation \textit{to all} $2\times2$ time-dependent Hamiltonians, a problem of current and central importance to quantum computing. 
%We demonstrate the use of the solution on a test model, that of a circularly polarised RF excitation in the laboratory frame. Indeed, the exact expression of $\mathsf{U}(t',t)$ is known thanks to a transformation into the rotating frame, yet none of the existing general purpose methods such as Magnus expansion or Floquet theory recovers it exactly. 
%
%In contrast, we show that path-sum arrives at the analytical solution. Second, 
As an example of application, we solve for the celebrated Bloch-Siegert dynamics %much more complex case 
of a linearly polarised RF excitation \emph{with no approximation} at all. The validity of the path-sum analytical solution is demonstrated over the entire driving range, % and for all types of driving (not necessarily harmonic). 
and physical interpretations for the various terms of the solution are provided.

%. It corresponds to the influence of the counter-rotating component of the RF field and the Bloch-Siegert effect. 
%Path-sum further leads to exact and compact representations of recently uncovered special solutions involved in two levels quantum dynamics, namely confluent Heun's functions otherwise known in general relativity and astrophysics \cite{Hortacsu2018}, see SI. 
%The two final examples are related to many-body Hamiltonians, including 
 We then show that path-sum is invariant against scale-transformations in the quantum state space, making it scalable to large quantum systems. Thanks to this, 
we consider $N$ like-spins coupled by the homonuclear dipolar coupling and spin diffusion under MAS. 
%In this later case, it is demonstrated that if the initial density matrix $\rho(0)$ is a pure state with a small number $k\ll N$ of up-spins, the evolution of $\rho(t)$ can be made analytically, even in the limit $N\to\infty$. 
The effects of MAS frequency and chemical shift offsets are illustrated \emph{analytically} on an organometallic molecule exhibiting 42 protons. 
%We emphasise the fact that from there, all corresponding simulations % presented below, 
%including the movies, typically took \emph{a minute} to be generated on a standard laptop. 
%More theoretical results related to spin chains are presented in the SI.  

\section{Quantum evolution and walks on graphs} 
Quantum systems with discrete degrees of freedom such as spin systems, obey a discrete analog to Feynman's path integrals. 
To illustrate this, define one history of a quantum system as a temporal succession of \textit{orthogonal} quantum states $h~:~|s_1\rangle\mapsto |s_2\rangle\mapsto |s_3\rangle\cdots$, each transition $|s_i\rangle\mapsto |s_{i+1}\rangle$ happening at a specified time $t_i$. Overall the history $h$ acquires a complex weight which is the product of the weights of all the transitions in the history. The weight of an individual transition $|s_i\rangle\mapsto |s_{i+1}\rangle$ is dictated by the Hamiltonian as $\langle s_{i+1} |\mathsf{H}(t_i)|s_{i}\rangle$.
 
A natural representation of such discrete histories is as walks on a graph. To see this, let $\G_t$ be the graph such that each vertex $v_i$ corresponds to one member $|s_i\rangle$ of an orthonormal basis for the entire state-space and give the directed edge $v_i\mapsto v_j$ the \emph{time-dependent} weight $\langle s_j |\mathsf{H}(t)|s_{i}\rangle$. 
In this picture, a system history as defined earlier is a walk on $\G_t$ and $\mathsf{H}(t)$ is the adjacency matrix of $\G_t$. Because the Hamiltonian is time-dependent, the graph itself is dynamical, see Fig~\ref{fig:PSstructure}(a,b) for an example. 

\begin{figure}
\begin{center}
\includegraphics[width=0.12\textwidth]{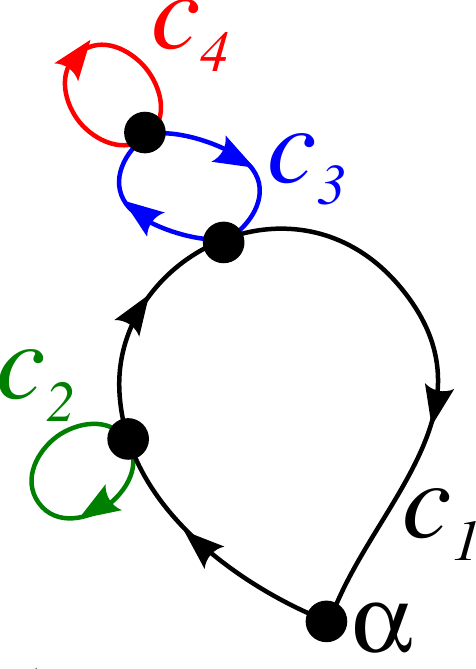}
  \end{center}
  \vspace{-4.5mm}
 \caption{\label{FigGraph}{\footnotesize Graph illustrating the use of path-sum.}}
  \vspace{-2mm}
\end{figure}

%\subsection*{General solution}
%\setlength{\columnsep}{8pt}%
%\begin{wrapfigure}[14]{c}{0.17\textwidth}
%  \begin{center}
%  \vspace{-5.5mm}
%    \includegraphics[width=0.12\textwidth]{Dessin1.pdf}
%  \end{center}
%  \vspace{-4.5mm}
%  \caption{\label{FigGraph}{\footnotesize Graph illustrating the use of path-sum.}}
%\end{wrapfigure}
Now just as for Feynman's path-integrals, the exact evolution of the system is obtained from the superposition of all its possible histories. Equivalently, every element $\langle s_j |\mathsf{U}(t)|s_{i}\rangle$ of the evolution operator $\mathsf{U}(t)$ is given by the sum over all walks from $v_i$ to $v_j$ on $\G_t$, including all possible jumping times for each transition between vertices. While individual walks are the discrete counterpart of Feynman diagrams, their algebraic structure is much better understood. Indeed, walks essentially behave as the natural integers \cite{Giscard2012}, in particular they can be uniquely factored  into products of prime walks: the simple cycles ($\mathcal{C}$) and paths ($\mathcal{P}$) of the graph which do not visit any vertex more than once. Since, by nesting simple cycles and paths into one another there is a unique way of reconstructing any walk, summing over all walks is achieved upon summing over all possible nestings of simple cycles and paths. For example, in a graph with a single simple cycle $c_1$, all closed walks from a vertex $\alpha$ to itself are of the form $c_1^n$, i.e. $c_1$ repeated $n$ times. Therefore the sum of all such walks is formally $\sum_n c_1^n = 1/(1-c_1)$ (Fig.~\ref{FigGraph}). 
In case another cycle $c_2$ is accessible to a walker while walking along $c_1$, then the sum of all walks will take on the form $1/\big(1-c_1/(1-c_2)\big)$. If instead, both $c_1$ and $c_2$ are immediately accessible from the starting point $\alpha$, the sum of all walks will be $1/(1-c_1 - c_2)$.
Finally, if two cycles $c_2$ and $c_3$ with different starting points are both accessible while walking along $c_1$, then the sum of all walks will be $1/\big(1-c_1/(1-c_2)\times 1/(1-c_3)\big)$. There is a unique way to combine these constructions to describe the sum of all walks with chosen starting and ending points on any graph. For example, the walks from $\alpha$ to itself on the graph of Fig.~\ref{FigGraph} formally sum up to 
$$
\sum_{w\,\text{walk:}\,\alpha\to\alpha}\!\!\!\! w\, = \frac{1}{1-c_1\frac{1}{1-\textcolor{DarkGreen}{c_2}}\frac{1}{1-\textcolor{blue}{c_3}\frac{1}{1-\textcolor{red}{c_4}}}}.
$$
This procedure yields any $\langle s_j |\mathsf{U}(t)|s_{i}\rangle$ as branched continued fractions comprising only the weights of the simple cycles and paths of the graph. See Fig~\ref{fig:PSstructure}(c,d,e). Because the graph is finite, there are finitely many such cycles and paths and the fraction is \emph{finite} in both depth and breadth. It is thus unconditionally convergent when calculated numerically. 
The same principles apply regardless of whether the Hamiltonian depends on time or not, in the former case however the theory relies on two-times functions $f(t',t)$ that multiply via a non-commutative convolution-like product 
\begin{equation}\label{DefStarProduct}
\big(f\ast g\big)(t',t) := \int_t^{t'}\!\!f(t',\tau)g(\tau,t)\, d\tau.
\end{equation}
This means that for general time-dependent Hamiltonians the continued fraction formulation for $\mathsf{U}(t)$ involves products and resolvent with respect to the $\ast$ multiplication and that the order of traversal of the edges along the cycles must be respected. 
The $\ast$-resolvents such as $(1_\ast-f)^{\ast-1}$ with $1_\ast \equiv \delta(t'-t)$ the Dirac delta distribution, are solutions to linear Volterra equations of the second kind. They concentrate most of the analytical complexity of the problem, rarely having a closed form expression in terms of algebraic mathematical functions. Yet, such $\ast$-resolvent can be always represented analytically by the super-exponentially convergent Neumann expansion \cite{Giscard2015} $1_\ast+\sum_{n>0} f^{\ast n}$.

\section{Two-level systems with time-dependent Hamiltonians}
\subsection{General solution}\label{Gen2}
\begin{figure*}[!t]
\begin{center}
\includegraphics[width=.9\textwidth]{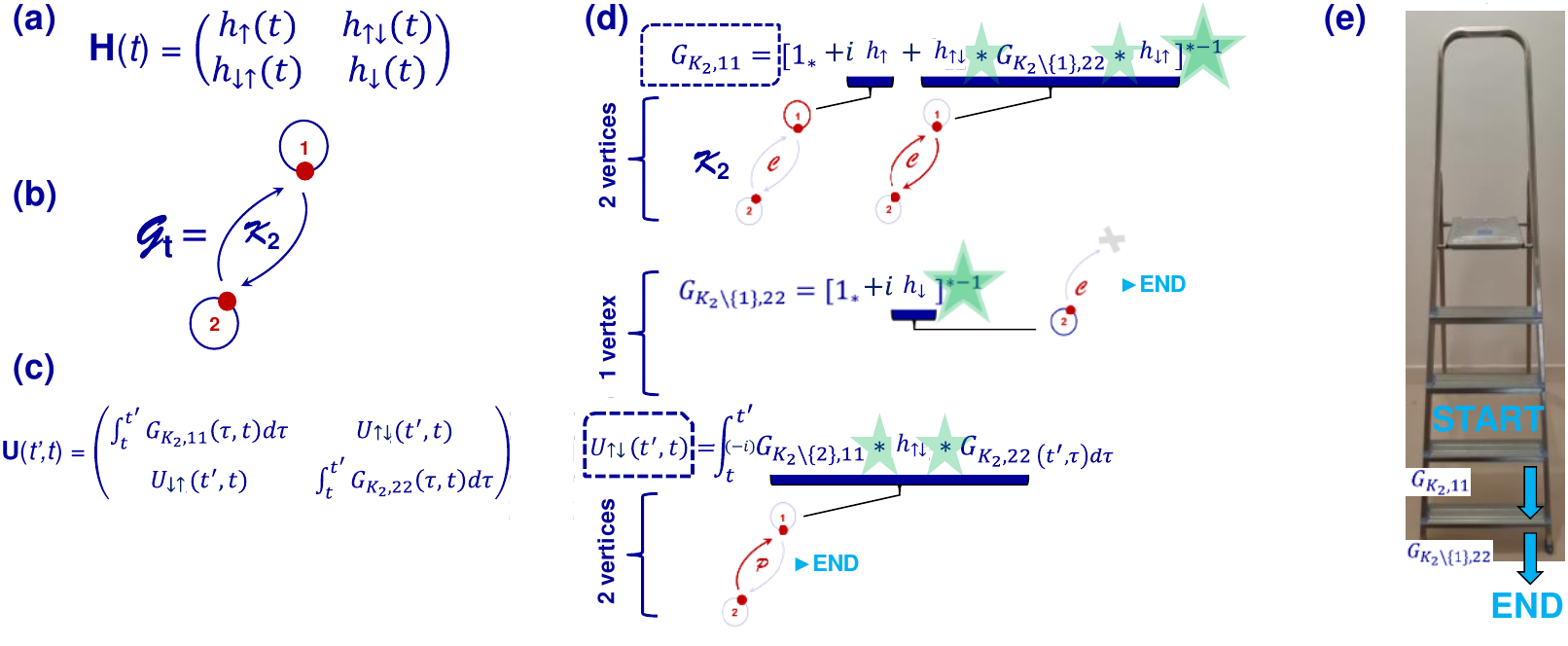}
\vspace{-3mm}
\caption{\textbf{The path-sum continued fraction} for the exact calculation of the entries of $\mathsf{U}(t',t)$ is always of finite depth and breadth. \textbf{(a)} %The
Illustrative example of a $2\times2$ time-dependent Hamiltonian $\mathsf{H}(t)$% which is 
, involved for instance in spin dynamics. \textbf{(b)} %The 
%Associated 
Dynamical graph $\mathcal{G}_t= K_2$ with adjacency matrix $\mathsf{H}(t)$. Circles correspond to self-loops (diagonal terms of $\mathsf{H}(t)$) and directed edges %correspond 
to off-diagonal terms. The associated weights are the entries of $\mathsf{H}(t)$. \textbf{(c)} Evolution operator $\mathsf{U}(t',t)$ as seen by path-sum, with integrals of the $G_{K_2,ii}$ quantities. \textbf{(d)} Step by step evaluation of $G_{K_2,11}(t',t)$ (dashed rectangle) showing the finite character of the % involved 
continued fraction. The sum is performed on the %finite 
simple cycles ($\mathcal{C}$) of length 1 and 2 %respectively 
(these 
%corresponding walks 
are indicated in red---other edges are indicated in dashed grey lines). At each step of the continued fraction, a vertex is removed (grey cross) and we work on subgraphs of less and less complexity. The calculation of entry $\mathsf{U}_{21}$ from $G_{K_2,11}$ is also illustrated, it includes a single term with two $\ast$-products as the graph has a single simple path ($\mathcal{P}$) from 2 to 1 (red arrow). 
\textbf{(e)} A pictorial representation of the "descending ladder principle". 
%for the path-sum continued fraction. 
The calculation starts at the top of the ladder with each $\ast$-inverse leading to the step below and ending in all
% (finite dimensional Hamiltonian) 
cases on the ground. For $2\times 2$ Hamiltonians there are only 2 steps on the ladder, i.e. all continued fractions stop at depth 2. For all $3\times 3$ and $4\times 4$ Hamiltonians the continued fractions stop at depth 3 and 4, respectively.}%: the continued fraction is of finite depth.}
\label{fig:PSstructure}
\vspace{-5mm}
\end{center}
\end{figure*}
Determining the dynamics of two-level systems driven by time-dependent Hamiltonians \emph{is still an open problem}, which continues to be a very active area of research \cite{Kayanuma1994, Blanes2009, Xie2010, Irish2005, Ashhab2007, Saiko2008, Gangopadhyay2010, Rapedius2015, Barnes2012, Yan2015, Schmidt2018}. This is because of the experimental relevance of such systems; their role as theoretical models; and the need to master the internal evolution of qubits undergoing quantum gates \cite{Barnes2012, Zeuch2018}. The most general two-level Hamiltonian is
\begin{equation}\label{HGen}
\mathsf{H}(t)=\begin{pmatrix}
h_{\up}(t)& h_{\up\down}(t)\\
 h_{\down\up}(t)&h_{\down}(t)
\end{pmatrix}.
\end{equation}
In this expression we only require that $h_{\down\up}(t)$, $h_{\up\down}(t)$, $h_{\up}(t)$ and $h_{\down}(t)$ be bounded functions of time over the interval $[t,t']$ of interest. So far, \emph{no analytical expression has been found} for the corresponding evolution operator $\mathsf{U}(t)$ defined as the unique solution of Eq.~(\ref{DefU}) with the Hamiltonian of Eq.~(\ref{HGen}). It is known that particular choices for $\mathsf{H}(t)$ lead to evolution operators that involve higher transcendantal functions \cite{Xie2010, Braak2011}. Thus the best possible result for the general $\mathsf{U}(t)$ is that each of its entries be described as solving a defining equation, and that an \emph{analytical} mean of generating this solution be presented.

This is exactly what path-sum achieves for all time-dependent two-level systems. Following the process of Fig.~(\ref{fig:PSstructure}) we get: 
\begin{align}\label{U22FORM}
&\mathsf{U}(t',t)_{\up\up}=\int_{t}^{t'}G_\up(\tau,t) d\tau,\quad \mathsf{U}(t',t)_{\down\down}=\int_{t}^{t'}G_\down(\tau,t) d\tau,\\
%&\mathsf{U}(t',t)_{12}=\int_{t}^{t'}\!\!\!\left(\delta(t'-\tau)+e^{-i\int_{\tau}^{t'} H_\up(\tau')d\tau'}\!\right)\!H_{\up\down}(\tau)\mathsf{U}(\tau,t)_{11} d\tau,\\
%&\mathsf{U}(t',t)_{21}=\int_{t}^{t'}\!\!\!\left(\delta(t'-\tau)+e^{-i\int_{\tau}^{t'} H_\up(\tau')d\tau'}\!\right)\!H_{\up\down}(\tau)\mathsf{U}(\tau,t)_{11} d\tau,\\
%\end{align*}
%Using the results for $\mathsf{U}(t',t)_{11}$ and $\mathsf{U}(t',t)_{22}$ the off-diagonal results further simplify to 
%\begin{align*}
&\mathsf{U}(t',t)_{\down\up}=\nonumber\\
&-i\int_{t}^{t'}\!\!\!\!\int_{t}^{\tau_0}\!\!\!\int_{\tau_1}^{\tau_0}\!\!\!\Big(\delta(\tau_0-\tau_2)-ih_{\down}(\tau_0)\,e^{-i\!\int_{\tau_2}^{\tau_0} h_{\down}(\tau_3)d\tau_3}\!\Big)\times\nonumber\\
&\hspace{40mm}\times h_{\down\up}(\tau_2)G_{\up}(\tau_1,t)\,\, d\tau_2\,d\tau_1\,d\tau_0,\nonumber\\
&\mathsf{U}(t',t)_{\up\down}=\nonumber\\
&-i\int_{t}^{t'}\!\!\!\!\int_{t}^{\tau_0}\!\!\!\int_{\tau_1}^{\tau_0}\!\!\!\Big(\delta(\tau_0-\tau_2)-ih_{\up}(\tau_0)\,e^{-i\!\int_{\tau_2}^{\tau_0} h_{\up}(\tau_3)d\tau_3}\!\Big)\times\nonumber\\
&\hspace{40mm}\times  h_{\up\down}(\tau_2)G_{\down}(\tau_1,t)\,\, d\tau_2\,d\tau_1\,d\tau_0.\nonumber
\end{align}
%&\mathsf{U}(t',t)_{\up\down}=-i\int_{t}^{t'}\!\!\!\!\int_{t}^{\tau_0}h_{\up\down}(\tau_0)G_{\down}(\tau_1,t)\, d\tau_1d\tau_0\\
%&+-\!\int_{t}^{t'}\!\!\!\!\int_{t}^{\tau_0}\!\!\!\int_{\tau_1}^{\tau_0}\!\!\!\!h_{\up}(\tau_0)\,e^{-i\!\int_{\tau_2}^{\tau_0}\!\! h_{\up}(\tau_3)d\tau_3}h_{\up\down}(\tau_2)G_{\down}(\tau_1,t)\, d\tau_2d\tau_1d\tau_0.
while the `usual' $\mathsf{U}(t)$ is actually $\mathsf{U}(t,0)$.
The two-times Green's functions $G_{\up}:=(1_\ast-K_\up)^{\ast-1}$ and $G_{\down}:=(1_\ast-K_\down)^{\ast-1}$ solve linear Volterra equations of the second kind, e.g. for $G_\up$ 
\begin{equation}
G_\up(t',t) = \delta(t',t)+\int_{t}^{t'}\!\! K_\up(t',\tau) \,G_\up(\tau,t)\,d\tau,\label{GEq}
\end{equation} 
and similarly for $G_\down$. The kernel $K_\up$ of the above equation is
\begin{align}\label{KernelExplicit}
K_\up(t',t)&=-ih_{\up}(t')\\
&\hspace{-10mm}-\!\!\int_{t}^{t'}\!\!\int_{\tau_1}^{t'}\!\!h_{\up \down}(t')\Big(\delta(\tau_2-\tau_1) -i h_\down(\tau_2)\, e^{-i\int_{\tau_1}^{\tau_2}h_\down(\tau_3)d\tau_3}\Big)\times\nonumber\\
&\hspace{47mm}\times h_{\down\up}(\tau_1) d\tau_2 d\tau_1. \nonumber
\end{align}
while kernel $K_\down$ entering $G_{\down}$ is obtained upon replacing up arrows by down arrows and vice-versa in Eq.~(\ref{KernelExplicit}). 

\begin{comment}
All of these results can be succintly presented using $\ast$-products as:
\begin{align}\label{U22general}
&\mathsf{U}(t',t) =-i\times\begin{pmatrix} 
i(1\ast G_{\up})&1\ast F_\up\ast h_{\up\down}\ast G_{\down}\\
1\ast F_\down\ast h_{\down\up}\ast G_{\up}&i(1\ast G_\down)
\end{pmatrix},
\end{align}
where 
\begin{align*}
F_\up&:=(1_\ast+h_\up e^{-i(1\ast h_\up)})\equiv \delta(t'-t)+h_\up(t')e^{-i\int_{t}^{t'}h_\up(\tau)d\tau},
\end{align*}
and similarly for $F_\down$. The Volterra equations for the two-times Green's functions $G_{\up}:=(1_\ast-K_\up)^{\ast-1}$ and $G_{\down}:=(1_\ast-K_\down)^{\ast-1}$ are then $G_\up = 1_\ast+K_\up \ast G_\up$ and $G_\down = 1_\ast+K_\down \ast G_\down$.
%, that is
%\begin{equation}
%G_\up(t',t) = \delta(t',t)+\int_{t}^{t'}\!\! K_\up(t',\tau) \,G_\up(\tau,t)\,d\tau,%\label{GEq}
%\end{equation} 
%where 
%$K_\up= -ih_{\up}-h_{\up \down}\ast F_\down \ast h_{\down\up}$ (see Eq.~(\ref{KernelExplicit}) for the explicit form) is called the kernel of the Volterra Eq.~(\ref{GEq}).we have $G_\up = 1_\ast+K_\up \ast G_\up$, that is
%It is a general feature of path-sums that the kernel functions appearing in the calculations are \emph{separable} (also called \textit{degenerate}), that is they can always be written as sums of products of functions of a single time variable $K(t',t)=\sum_{i=1}^q A_i(t')B_i(t)$. This property holds true for general $N$-body Hamiltonians and facilitates the analysis of the Volterra equations from path-sums \cite{Pleshchinskii1995}. 
\end{comment}

Should a closed form expression for the solution the Volterra equation be out of reach--e.g. because it is a transcendental function as is typically the case 
\cite{Xie2010}--the solution is at least analytically available from its Neumann expansion; in the case Eq.~(\ref{GEq}) it is $G_\up=1_\ast+\sum_{n>0} K_\up^{\ast n}$. 
If every entry of the Hamiltonian is a bounded function of time, this series representation converges super-exponentially and uniformly \cite{Giscard2015}. 
Alternatively, Volterra equations can also be solved numerically \cite{Hackbusch1995}.
\subsection{Bloch-Siegert dynamics}
\subsubsection{Background}
The Bloch-Siegert Hamiltonian, here denoted $\mathsf{H}_{BS}(t)$, 
is possibly the simplest model to exhibit non-trivial physical effects due to time-dependencies in the driving radio-frequency fields. 
The detailed study of these effects is of paramount importance in the broad field of quantum computing, as they have a deleterious impact on qubit driving and stored quantum information \cite{Zhang2018}. 
The Hamiltonian reads
\begin{equation}\label{HBS1}
\mathsf{H}_{BS}(t)=\begin{pmatrix}\omega_0/2& 2 \beta \cos (\omega t)\\
 2 \beta \cos (\omega t)& -\omega_0/2
\end{pmatrix}.
\end{equation}
In these expressions, the coupling parameter $\beta$ is the amplitude of the radio-frequency field.

Continuing research over the last decades has produced perturbative expressions for the Bloch-Siegert shifts and evolution operator, starting from Shirley's seminal work \cite{Shirley1965}.
%Yet, there is still no general analytical approach to determine the Bloch-Siegert evolution operator exactly, and all existing approaches are limited one way or another to low order expansions. 
Beyond the rotating wave approximation--which omits the field's counter-rotating terms and is limited to near resonant $\omega\sim\omega_0$ ultra-weak couplings $\beta/\omega\ll 1$--one of the most successful approaches  used a combination of Floquet formalism and almost degenerate perturbation theory \cite{Aravind1984}. Still, this could only approximate the temporal dynamics in the vicinity of resonance in the weak coupling regime $\beta/\omega\lesssim 0.6$. 

In the case of quantum systems driven by large amplitude fields $\beta/\omega>1$ to $\beta/\omega\gg 1$ \cite{Ashhab2007}, these approaches are no longer sufficient. Yet, such systems are of current fundamental interest, as short associated electromagnetic pulses can manipulate qubits on a large bandwidth. Recently, several methods have thus been designed to overcome the limitations of the theoretical treatment \cite{Lu2012,Yan2015,Zhang2015,Yan2017}. These are based on various unitary transformations leading to approximate analytical expressions over an extended driving range. Although these methods are clearly beyond perturbation theory and what Floquet formalism can realistically achieve, they still neglect terms corresponding to multi-photon transitions. Although not dominant, these terms lead to real features in the true Bloch-Siegert dynamics that are as yet unaccounted for \cite{Lu2012}. These are visible in qubit driving and time-optimal quantum control experiments, for which determining the Bloch-Siegert dynamics exactly is thus still full of promises \cite{Laucht2016}.
%Finally, the methods employed are specialised to the Bloch-Siegert Hamiltonian and cannot easily be made to apply to other situations. 
In spite of the theoretical efforts,
\emph{a non-perturbative truly analytical solution at all orders over the entire coupling range, on and off resonance}, is ultimately lacking.
% We show that path-sum not only provides such an exact mathematical treatment, but also that it does so for all time-dependent Hamiltonians.

\subsubsection{Path-sum solution}\label{BSPS}
Although this is not required by the path-sum method, we pass in the interaction picture to alleviate the notation, yielding the Bloch-Siegert Hamiltonian as
\begin{equation}\label{HBS}
\mathsf{H}_{BS}(t)=2 \beta \cos (\omega t) \cos (\omega_0 t)\sigma_x- 2 \beta \cos (\omega t) \sin (\omega_0 t)\sigma_y.
\end{equation}
Since in the rotating frame, $h_\up(t)=h_\down(t)=0$, the graph $K_2$ of Fig.~(\ref{fig:PSstructure}) has no self-loops and Eqs~(\ref{U22FORM}--\ref{GEq}) thus  give
\begin{figure*}[!t]
\begin{center}
\includegraphics[width=1\textwidth]{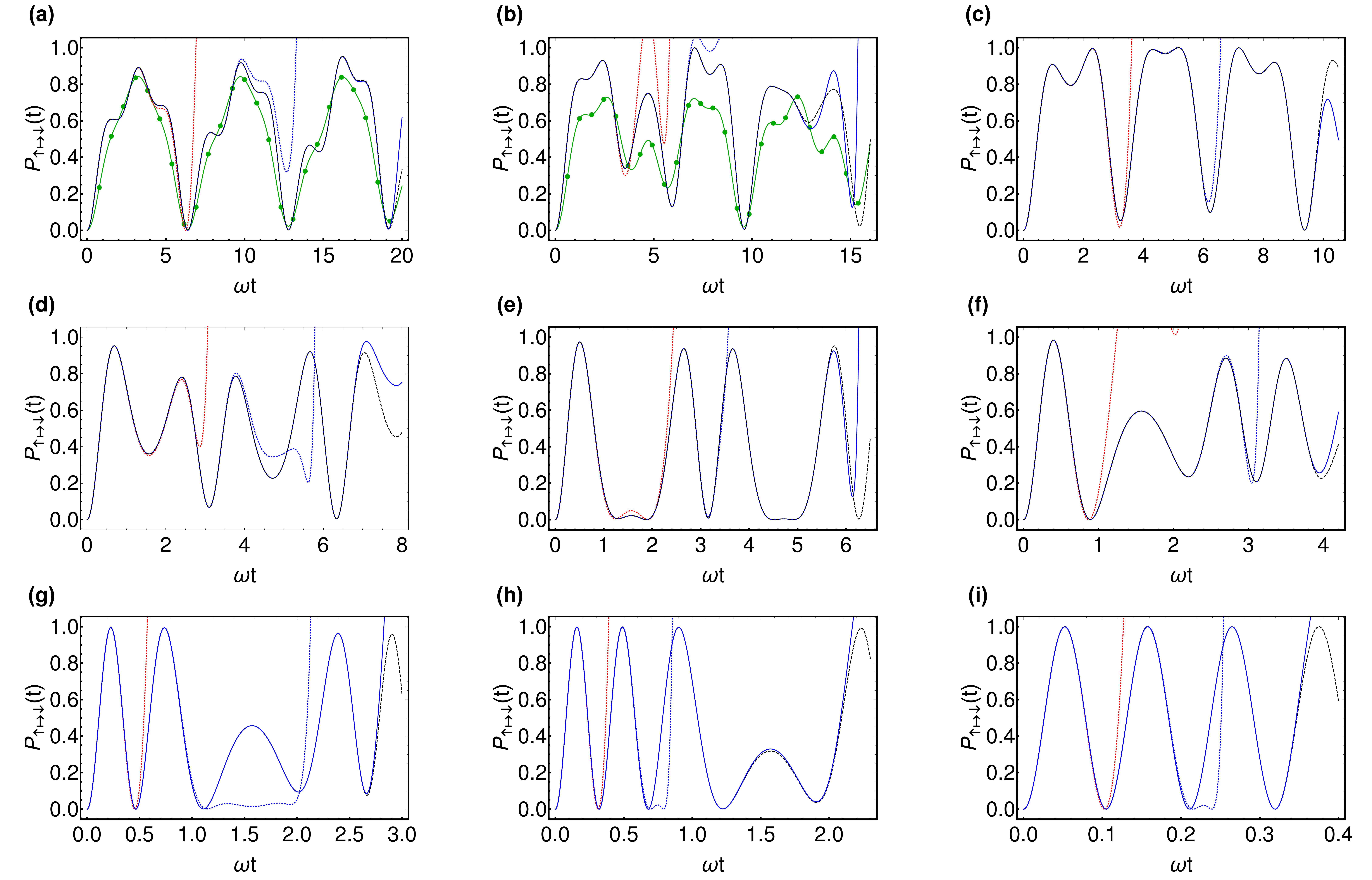}
\caption{\textbf{Bloch-Siegert dynamics:} resonant $\omega_0=\omega$ transition probability $P_{\up\mapsto \down}(t)$ as a function of time in the weak to ultra-strong coupling regimes: with 
\textbf{(a)} $\beta/\omega=0.5$, 
\textbf{(b)} $\beta/\omega=0.7$, 
\textbf{(c)} $\beta/\omega=0.9$,
\textbf{(d)} $\beta/\omega=1.2$, 
\textbf{(e)} $\beta/\omega=1.6$, 
\textbf{(f)} $\beta/\omega=2$,
\textbf{(g)} $\beta/\omega=3.5$, 
\textbf{(h)} $\beta/\omega=5$ and \textbf{(i)} $\beta/\omega=15$. Shown here are the numerical solution (dashed black line) and the fully analytical formulas for the Neumann expansions of the exact path-sum solution $P^{(3)}_{\up\to\down}(t)$ (dotted red line), $P^{(7)}_{\up\to\down}(t)$ (dotted blue line) and $P^{(13)}_{\up\to\down}(t)$ (solid blue line), see Appendix~\ref{2LevelExplicit}. As seen here, each of these formulas are equally valid throughout the coupling regimes, from weak to ultra-strong. Whenever longer times are desired, higher orders of the path-sum solution are readily available \emph{analytically}.   
Also shown in figures $\textbf{(a)}$ and $\textbf{(b)}$
are the second order Floquet theory \cite{Shirley1965} (solid green line and green points). The Floquet result is not shown in subsequent figures, where it is wildly inaccurate. Parameters : two level system driven by the Bloch-Siegert Hamiltonian of Eq.~(\ref{HBS}) \cite{Bloch1940,Shirley1965} starting in the $|\!\up\rangle$ state at $t=0$.}
\label{fig:PBlochSiegert}
\end{center}
\end{figure*}
\begin{align}
&\mathsf{U}(t)_{\up\up}=\int_{0}^{t}G_\up(\tau,0) d\tau,\quad \mathsf{U}(t)_{\down\down}=\int_{0}^{t}G_\down(\tau,0) d\tau,\nonumber\\
%&\mathsf{U}(t',t)_{12}=\int_{t}^{t'}\!\!\!\left(\delta(t'-\tau)+e^{-i\int_{\tau}^{t'} H_\up(\tau')d\tau'}\!\right)\!H_{\up\down}(\tau)\mathsf{U}(\tau,t)_{11} d\tau,\\
%&\mathsf{U}(t',t)_{21}=\int_{t}^{t'}\!\!\!\left(\delta(t'-\tau)+e^{-i\int_{\tau}^{t'} H_\up(\tau')d\tau'}\!\right)\!H_{\up\down}(\tau)\mathsf{U}(\tau,t)_{11} d\tau,\\
%\end{align*}
%Using the results for $\mathsf{U}(t',t)_{11}$ and $\mathsf{U}(t',t)_{22}$ the off-diagonal results further simplify to 
%\begin{align*}
&\mathsf{U}(t)_{\down\up}=\label{Udu}\\
&\hspace{5mm}-2i\beta\int_{0}^{t}\int_{0}^{\tau_1} \cos(\omega\tau_1)e^{i\omega_0 \tau_1}G_\up(\tau_0,0)d\tau_0d\tau_1,\nonumber\\
&\mathsf{U}(t)_{\up\down}=\nonumber\\
&\hspace{5mm}-2i\beta\int_{0}^{t}\int_{0}^{\tau_1} \cos(\omega\tau_1)e^{-i\omega_0 \tau_1}G_\down(\tau_0,0)d\tau_0d\tau_1,\nonumber
\end{align}
while 
$G_\up(t',t)=(1_\ast-K_\up)^{\ast-1}$, $G_{\down}=\big(1_\ast-K_\down)^{\ast-1}$ with 
\begin{align*}
K_\up(t',t) &=\frac{4 \beta^2}{\omega^2-\omega_0^2} \cos (\omega t') \left(k_{\up}(t) e^{-i \omega_0 (t'-t)}-k_{\up}(t')\right),\\
K_\down(t',t) &=\\&\hspace{-7mm}\frac{i \beta^2}{\omega^2-\omega_0^2} \left(1+e^{-2 i \omega t'}\right)\left(k_\down(t')-k_\down(t) e^{i (\omega +\omega_0)(t'-t)}\right),
\end{align*}
where $k_{\up}(t)=i \omega_0 \cos ( \omega t )+\omega  \sin (\omega t)$ and $k_\down(t)=e^{2 i \omega t } (\omega +\omega_0)-(\omega -\omega_0)$. In spite of the apparent divergences in the resonant case $\omega_0\to \omega$, the kernels $K_{\up}$ and $K_{\down}$ are actually well defined in this limit where they simplify to
\begin{align*}
K_\up(t',t)&=\\
&\hspace{-5mm}\frac{\beta^2}{\omega}  \left(i e^{2 i \omega t' }-i e^{2 i \omega t }-2 \omega  (t'-t)\right)e^{-i \omega t' } \cos (\omega t' ),\\
\intertext{and}
K_\down(t',t)&=\\
&\hspace{-10mm}\frac{\beta^2}{\omega } \left(-i+i e^{2 i (t'-t) \omega }+2 \omega  (t-t') e^{2 i \omega  t'}\right) e^{-i\omega t'}\cos ( \omega t' ).
\end{align*}
The peculiar mathematical nature of the resonant limit $\omega_0\to \omega$ is responsible for the apparence of the term $2\omega(t'-t)$ which is proportional to time in the kernel.\\[-.5em]

The quantity $G_\up$ as obtained from $K_\up$ has no closed form, rather it is a hitherto unknown higher special function. It is nonetheless analytically available thanks to the Neumann expansion $G_\up = \delta+\sum_k K_\up^{\ast k}=\delta +K_\up+K_\up\ast K_\up+\cdots$, which is \emph{unconditionally convergent} \cite{Giscard2015}. This observation holds for all $N\times N$ time-dependent Hamiltonians treated by path-sum.

\begin{figure*}[!t]
\begin{center}
\includegraphics[width=1\textwidth]{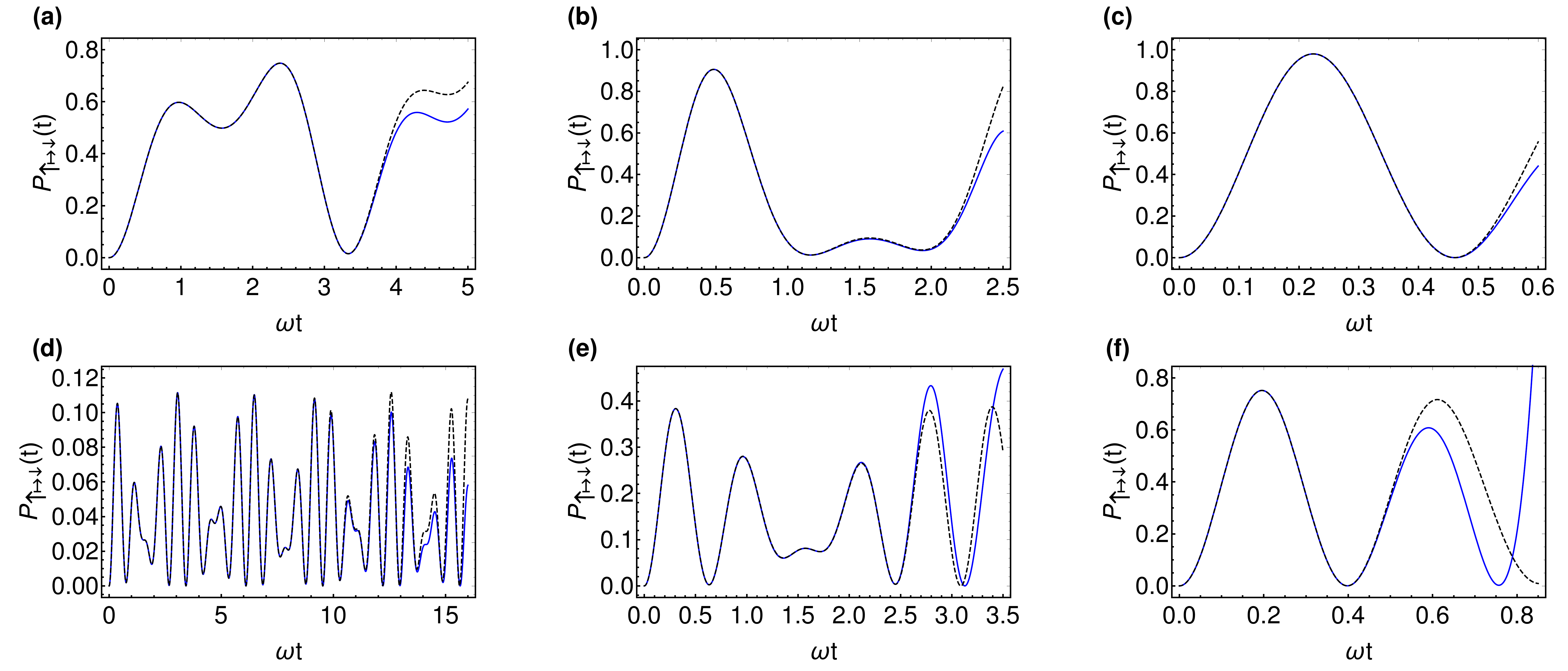}
\caption{\textbf{Bloch-Siegert dynamics:} off-resonance $\omega_0\neq \omega$ transition probability $P_{\up\mapsto \down}(t)$ as a function of time in the weak to strong coupling regimes: with 
\textbf{(a)} $\omega_0=2\omega$, $\beta/\omega=0.7$, 
\textbf{(b)} $\omega_0=2\omega$, $\beta/\omega=1.6$, 
\textbf{(c)} $\omega_0=2\omega$, $\beta/\omega=3.5$, 
\textbf{(d)} $\omega_0=8\omega$, $\beta/\omega=0.7$, 
\textbf{(e)} $\omega_0=8\omega$, $\beta/\omega=1.6$, and
\textbf{(f)} $\omega_0=8\omega$, $\beta/\omega=3.5$. 
Shown here are the numerical solution (dashed black line) and the fully analytical formula for the Neumann expansion of the exact path-sum solution at the fourth order $P^{(4)}_{\up\to\down}(t)$ (solid blue line). Parameters : two level system driven by the Bloch-Siegert Hamiltonian of Eq.~(\ref{HBS}) \cite{Bloch1940,Shirley1965} starting in the $|\!\up\rangle$ state at $t=0$.}
\label{fig:PBlochSiegertoff}
\end{center}
\end{figure*}
The Neumann expansion is well suited to analytical computations: observe that at order $n$ of this series, $G^{(n)}_\up=\delta+\sum_{k=1}^n K^{\ast k}$ is simply given as
$$
G_\up^{(n)}(t,0)=\int_{0}^{t}K_\up(t,\tau)G_\up^{(n-1)}(\tau,0)d\tau.
$$
Equivalently, it is sufficient to integrate the last term of the series at order $n-1$, namely $K_\up^{\ast n-1}$, to get $G_\up^{(n)}$: 
$$
G^{(n)}_\up(t,0)=\int_{0}^tK_\up(t,\tau)K_\up^{\ast n-1}(\tau,0)d\tau+G^{(n-1)}(t,0).
$$
These integrals are all analytically available and easily accessible: we reached order 13 in a minute on an ordinary laptop treating all parameters as formal variables \footnote{The \textsc{Mathematica} notebook generating these calculations is available for download at \url{http://www-lmpa.univ-littoral.fr/~plgiscard/}}. Vastly faster computations are achieved upon assigning parameter values before performing the integrals. 
This calculations give (here displaying the first two orders on resonance $\omega_0=\omega$),
\begin{align*}
G_{\up}(t,0) = \delta(t)&-\frac{\beta^2}{\omega }e^{-i\omega t}\cos (\omega t ) \left( -i e^{2 i \omega t }+2 \omega t+i\right) \\
&+\frac{\beta^4}{24 \omega ^3} e^{-3 i \omega t }\cos (
   \omega t ) \Big(3 i e^{6 i  \omega t }\\
   &\hspace{5mm}+6 e^{4 i \omega t } (-2i \omega^2 t^2  +2  \omega  t+i)\\
   &\hspace{5mm}+e^{2 i \omega t } (  8\omega^3 t^3 +12i\omega^2 t^2+12\omega t -15 i)\\
   &\hspace{5mm}-12  \omega t+6 i\Big)\\
&+\cdots
\end{align*}

Of particular interest for qubit-driving experiments is the evolution of the transition probability $P_{\up \mapsto \down}(t):=|U_{\down\up}(t)|^2$ between the two levels \cite{Angelo2005,Lu2012,Zeuch2018}. This quantity is usually found perturbatively using Floquet theory \cite{Shirley1965} as Magnus series again suffer from divergences \cite{Maricq1987}.
%The result of Eq.~[\ref{U22general}]
% yields the probability amplitude $A_{\up \mapsto \down}(t)$ as the solution of the Volterra equation  
%$A_{\up \mapsto \down} = F+K\ast A_{\up \mapsto \down}$ 
%where the function $F$ and kernel $K$ are given \ref{2LevelExplicit}. 
It is here easily accessible--$\mathsf{U}_{\down\up}$ being given by Eq.~(\ref{Udu}). We find that $P_{\up\to\down}(t)$ takes on the form of a Fourier-like series
\begin{equation}
P_{\up\to\down}(t)=\sum_{k=0}^\infty \sin(2k\omega t) S_{2k}(\beta,t)+ \cos(2k\omega t) C_{2k}(\beta,t),\label{Pseries}
\end{equation}
with $S_{2k}$ and $C_{2k}$ functions of $\beta$ and $t$, a representation of which is analytically available (see Appendix~\ref{2LevelExplicit}). This form of $P_{\up\to\down}(t)$ is due to the path-sum integral of Eq.~(\ref{Udu}), which resembles a Fourier transform. We emphasize that this is not a general feature of path-sum nor of $2\times2$ Hamiltonians, but solely of the present  Hamiltonian with linearly polarized driving. 

\subsubsection{Visualizing the solution}
Calculating $G_\up$ up to a finite order $n$ as indicated earlier $G_\up\equiv G_\up^{(n)}$, yields an expression $
P^{(n)}_{\up\to\down}(t)$ which includes all terms of Eq.~(\ref{Pseries}) up to $\sin\big((4n+2)\omega t\big)$ and $\cos\big((4n+2)\omega t\big)$, while  $S^{(n)}_{2k\leq 4n+2}$ and $C^{(n)}_{2k\leq 4n+2}$ are polynomials in $\beta$ and $t$ including up to $\beta^{4n+2}$ and $t^{4n+3-2k}$ and $t^{4n+2-2k}$, respectively. Finally, we found analytically that at all orders $n\geq0$, $P^{(n)}_{\up\to\down}(0)=0$ as expected, although this is non-trivial to check. For $t$ large enough $P^{(n)}_{\up\to\down}(t)$ may diverge: truncated path-sums are not necessarily unitary.

We plot on Fig.~\ref{fig:PBlochSiegert} the transition probabilities $P^{(3)}_{\up\to\down}(t)$, $P^{(7)}_{\up\to\down}(t)$ and $P^{(13)}_{\up\to\down}(t)$ as calculated analytically from the third, seventh and thirteenth orders of the Neumann expansion of the exact path-sum solution from the weak to the ultra-strong coupling regimes and always in the resonant case $\omega_0=\omega$. Here this situation was chosen because: i) it is mathematically the most difficult to approach exactly owing to the peculiar form of $K_\up$ which slows down convergence; and ii) it yields `compact' expressions more suitable for a `concise' presentation (Appendix~\ref{2LevelExplicit}). Higher order terms of the Neumann expansion are readily and analytically available, enabling precise evaluation of $P_{\up \mapsto \down}(t)$ up to any desired target time. Recall that, as discussed above, $P^{(13)}_{\up\to\down}(t)$ is actually a single analytical formula involving all even frequencies sines and cosines up to $\sin(54\omega t)$ and $\cos(54\omega t)$ with coefficients up to $\beta^{54}$. We stress here that the Neumann expansion of the path-sum solution is profoundly different from a Taylor series representation, as is e.g. manifest even at order 0, see Eq.~(\ref{P0OSC}) below and Appendix.~(\ref{2LevelExplicit})
%We know that, at worst $|P_{\up\to\down}(t)-P^{(n)}_{\up\to\down}(t)|\leq t^{n+1}/(n+1)!$.

%We plot on Fig.~\ref{fig:PBlochSiegert} the transition probability $P_{\up \mapsto \down}(t)$ as calculated analytically from the second and third orders of the Neumann series $A_{\up \mapsto \down}=F\ast(1_\ast+\sum_{n>0}K^{\ast n})$ in the resonant case $\omega_0=\omega$. This situation was chosen because: i) it is mathematically the most difficult to approach exactly (see \ref{BSAppendix}); and ii) it yields compact expressions more suitable for a concise presentation. Higher order terms of the Neumann series are readily and analytically available, enabling precise evaluation of $P_{\up \mapsto \down}(t)$ at any desired target time.
The fact that the same expression for $P^{(n)}_{\up \mapsto \down}(t)$ is an equally good approximation to the exact transition probability in all parameter regimes, i.e. from $\beta/\omega_0\ll1$ to $\beta/\omega_0\gg1$ is a signature that the path-sum approach is non-perturbative. For the same reason, we observe that $P^{(n)}_{\up \mapsto \down}(t)$ captures roughly the same number of spin flips in time regardless of $\beta$: empirically order 3 reproduces 1--2 flips, order 7 gets 2--3 flips, order 13 captures 4--5.  
%See \ref{BSAppendix} for complete calculations related to the Bloch-Siegert Hamiltonian.
At the opposite, Floquet theory, which is inherently perturbative, only works for $\beta/\omega\ll1$ \cite{Shirley1965}, while the diverging Magnus series is limited to very short times.

In Fig.~(\ref{fig:PBlochSiegertoff}) we show the off-resonance $\omega_0\neq \omega$ dynamics of the analytical transition probability $P^{(4)}_{\up\to\down}(t)$ obtained from the fourth order Neumann expansion. Irrespectively of the coupling strength, at any fixed finite order $n$, $P^{(n)}_{\up\to\down}(t)$ is reliable for longer times as we get farther from resonance, for which convergence of the Neumann expansion is slowed by the presence of a linear term in $K_\up$. Once again, this is purely a feature of the Bloch-Siegert Hamiltonian and not of the path-sum approach. 

\subsubsection{Physical insights}
Now that we have analytical formulas for the transition probability without the rotating wave approximation, we may gain novel insights into the Bloch-Siegert dynamics. For example, we can calculate the spin-flip duration $t_{sf}$, i.e. the time at which $P_{\up\to\down}(t)$ first peaks close to 1 when on resonance $\omega_0=\omega$. Analysis of Eq.~(\ref{Pseries}) with e.g. the analytic expressions of Appendix~\ref{2LevelExplicit} shows that $C_0(\beta,t)$ is the dominant contribution to $t_{sf}$ in the weak coupling regimes $\beta/\omega\lesssim1/2$, while the $C_{2k>0}$ and $S_{2k}$ functions describe further oscillations smaller by a factor of at least $\beta^2$. Extracting $t_{sf}$ from $C_0$ leads to 
\begin{align}
t_{sf}&=\label{eqtpk}\\
&\hspace{-5mm}\frac{1}{2\sqrt{2}}\sqrt{\frac{12}{\beta^2}-\frac{15}{\omega^2}+\frac{\sqrt{3}}{\beta^4\omega^2}\sqrt{91 \beta^8-88\beta^6\omega^2+16\beta^4\omega^4}}\nonumber\\
&=\frac{1}{\beta}\sqrt{\frac{1}{2} \left(3+\sqrt{3}\right)}-\frac{\beta}{8\omega^2}\sqrt{\frac{1}{2} \left(129+67 \sqrt{3}\right)}\nonumber\\
&\hspace{5mm}-\frac{\beta ^3}{128 \omega^4}\sqrt{\frac{1}{2} \left(16131+5545 \sqrt{3}\right)} +O\left(\beta ^4\right).\nonumber
\end{align}
\begin{figure}[!t]
\begin{center}
\includegraphics[width=.48\textwidth]{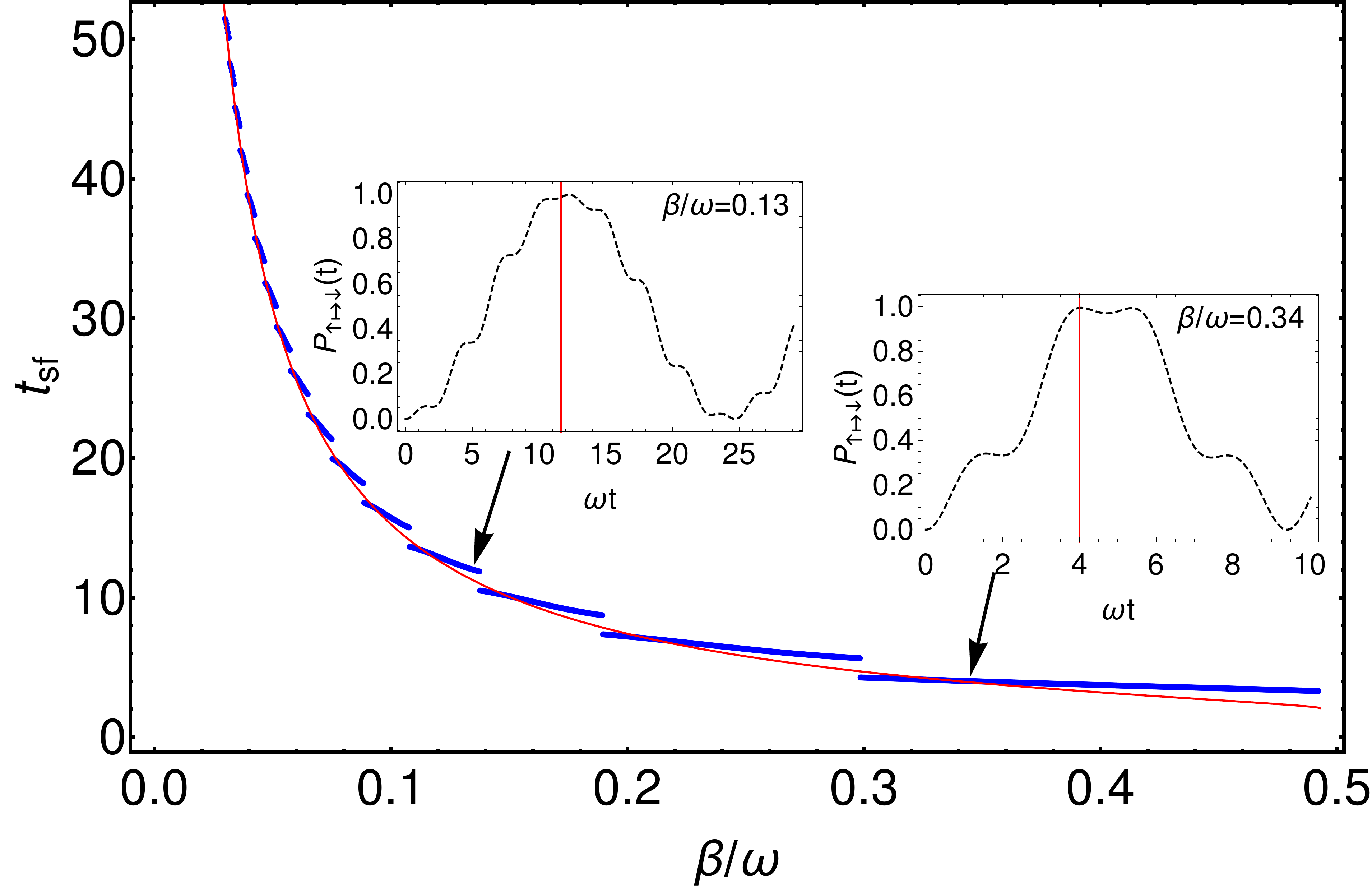}
\caption{\textbf{Bloch-Siegert dynamics:} Resonant $\omega_0=\omega$ spin-flip duration $t_{sf}$, at which $P_{\up\to\down}(t)$ first peaks at or close to 1, as a function of the coupling strength $\beta$. Shown here are the exact formula of Eq.~(\ref{eqtpk}) (solid red line) and fully numerical results (blue dots). Discontinuities in the numerical results are due to small oscillations of $P_{\up\to\down}(t)$, which make $t_{sf}$ undergo discrete jumps as one wins over the others. These are well captured analytically by a more advanced analysis including the $S_{2k}$ and $C_{2k>0}$ functions. In insets: two examples of time evolution of $P_{\up\to\down}(t)$, the straight red lines are the predictions of Eq.~(\ref{eqtpk}).}
\label{fig:tpk}
\end{center}
\end{figure}
This is remarkably close to the results obtain from numerical calculations, see Fig.~(\ref{fig:tpk}).
Mathematically, Eq.~(\ref{eqtpk}) assumes $\beta/\omega<2 \sqrt{\frac{1}{91} \left(11-\sqrt{30}\right)}\simeq 0.49$. Beyond this point the above estimate yields a complex number as the real solution switches to another root of the derivative of $C_0$.   

Even better analytical formulas for $t_{sf}$ with domains of validity that go much further into the stronger coupling regimes and accurately reflect its discrete jumps are immediately available, however they cannot be expressed in terms of radicals anymore and are not reproduced here owing to length concerns.\\[-.5em] 

\begin{figure}[!t]
\begin{center}
\includegraphics[width=.48\textwidth]{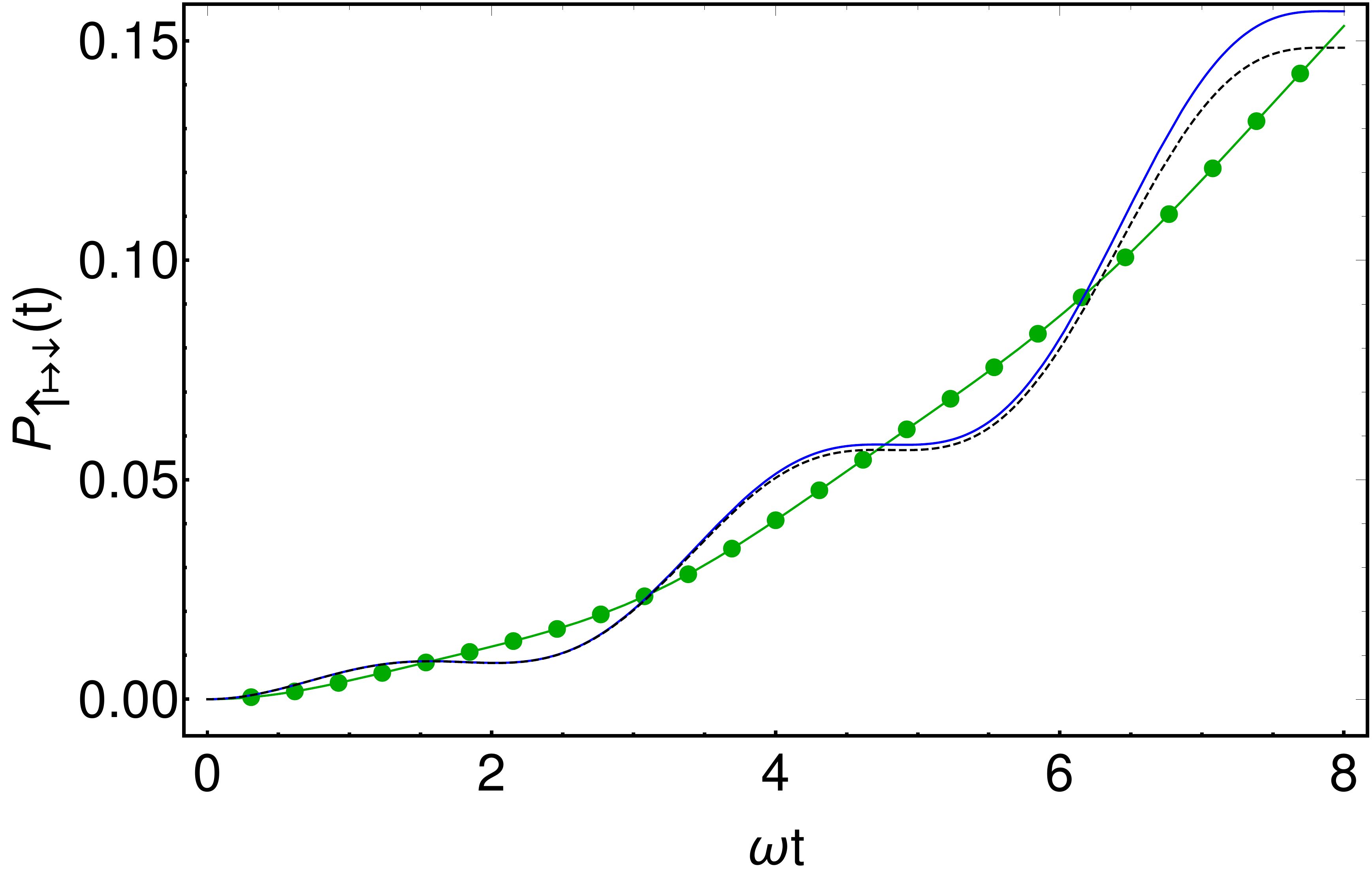}
\caption{\textbf{Bloch-Siegert dynamics:} resonant $\omega_0=\omega$ transition probability $P_{\up\to\down}(t)$ in the ultra-weak coupling regime $\beta/\omega=0.05$ for short times. Shown here are the order 0 formula $P^{(0)}_{\up\to\down}(t)$ of Eq.~(\ref{P0OSC}) (solid blue line), second order Floquet theory \cite{Shirley1965} (solid green line and green points) and the numerical solution (dashed black line).}
\label{fig:P0OSC}
\end{center}
\end{figure}

Also of interest are the changes affecting the dynamics of the transition probability $P_{\up\to\down}(t)$ as $\beta/\omega$ is increased from the weak to strong regimes. For example, in the ultra-weak coupling regime $\beta/\omega\ll 1$, the path-sum solution reproduces small oscillations around the Floquet calculations which are present in the  numerical solution, see Fig.~(\ref{fig:P0OSC}). In fact, these small oscillations are already captured by the order 0 of the Neumann expansion of the path-sum solution (!), for which $G_\up^{(0)}=\delta(t',t)$ and 
\begin{equation}\label{P0OSC}
P^{(0)}_{\up\to\down}(t)=\frac{\beta^2 t}{\omega}\sin(2\omega t)+\frac{\beta ^2}{2 \omega ^2}+\beta ^2 t^2-\frac{\beta ^2 }{2 \omega ^2}\cos(2\omega t).
\end{equation}
This shows that the oscillations missed by earlier treatments have a linearly-growing amplitude at short times on the order of $\beta^2 t$, originate purely from the counter-rotating terms,  and never trully vanish as long as $\beta\neq 0$. The diverging parabola in $\beta^2 t^2$ reflects the humble beginning of the Rabi oscillation, unsurprisingly missed by order 0. As $\beta$ is increased the small oscillations compete with the background Rabi oscillations, thereby giving rise to intricate intermediary effects seen in Fig.~(\ref{fig:PBlochSiegert}). This competition also explains why $P_{\up\to\down}(t)$ does not always peak at 1, as it results from a complicated superposition of oscilatory terms, in agreement with Eq.~(\ref{Pseries}).\\[-.5em] 

We conclude the discussion on physical insights into the Bloch-Siegert dynamics by studying Coherent Destruction of Tunneling (CDT) \cite{grifoni1998} in the strong coupling $\beta/\omega_0 \gg 1$. This situation is well suited to the use of a general property of Neumann series that allows for arbitrary accelerations of their convergence in the presence of dominant terms  \cite{VolterraGiscard}. Note that this procedure is always available when expanding path-sum solutions, and is thus not specific to the Bloch-Siegert Hamiltonian.

\begin{figure*}[!t]
\begin{center}
\includegraphics[width=1\textwidth]{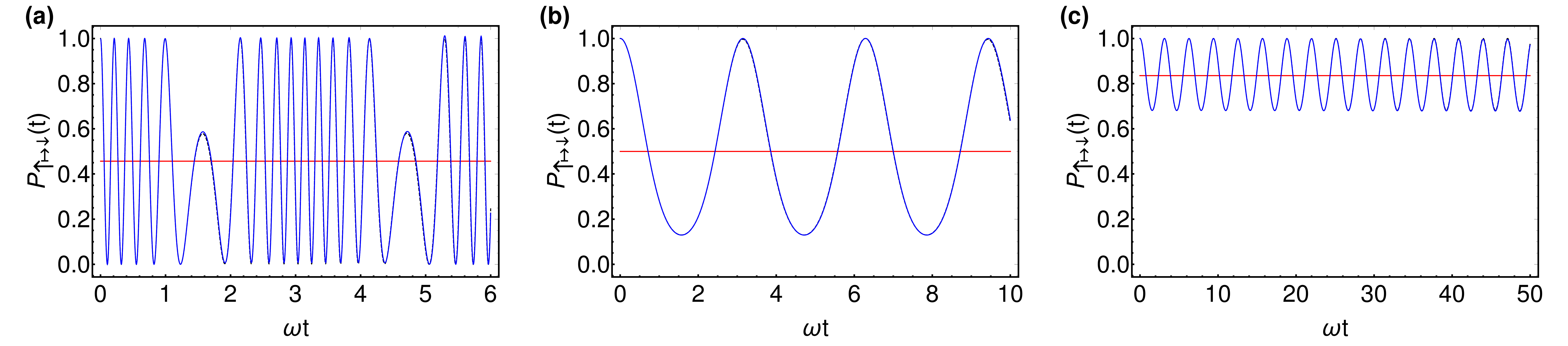}\\
\includegraphics[width=1\textwidth]{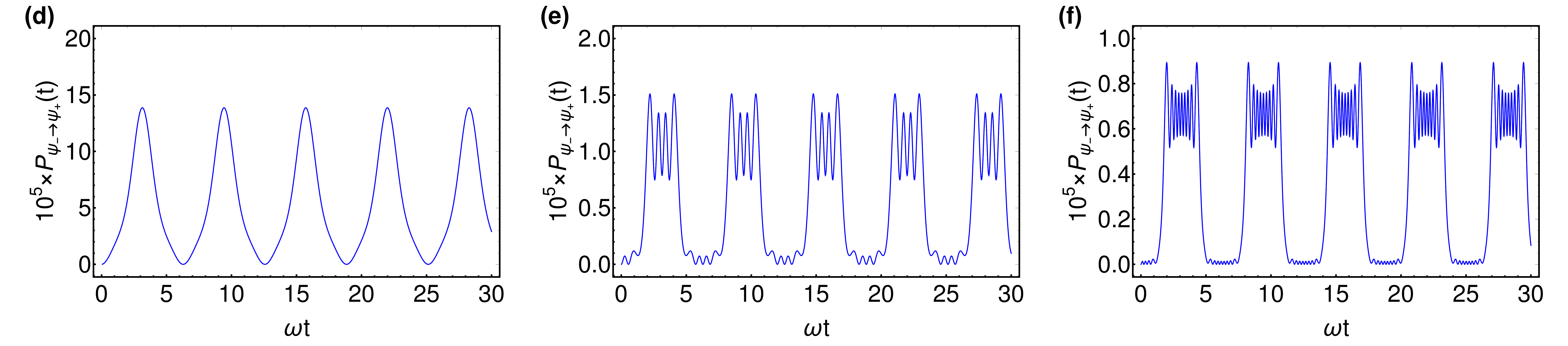}\\
\includegraphics[width=1\textwidth]{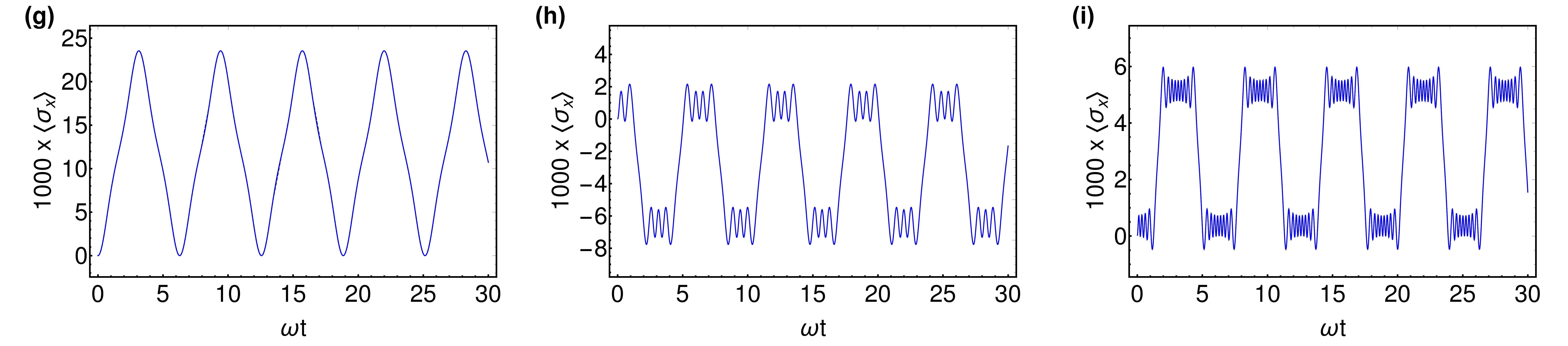}
\vspace{-3mm}
\caption{\textbf{Coherent destruction of tunneling:} Top line, return probability $P_{\up\to\up}(t)$ in the ultra-strong coupling regime $\beta/\omega_0=30$ for \textbf{(a)} $\omega=4\omega_0$; \textbf{(b}) $\omega=20\omega_0$; and \textbf{(c}) $\omega=100\omega_0$. Shown here are $P^{(acc,0)}_{\up\to\up}(t)$ as given by Eq.~(\ref{ReturnProb}) (solid blue line), 
the numerical solution (dashed black line), and its predicted time-average Eq.~(\ref{TimeAverage}) (solid red straight line, indistinguishable from the numerically computed time-average). Middle line: transition probability $P_{\psi_-\to\psi_+}(t)$ for a system starting in the $|\psi_-\rangle$ state at $t=0$ with: \textbf{(d)} $4\beta/\omega=2.404...$, first zero of $J_0(4\beta/\omega)$; \textbf{(e)} $4\beta/= 11.79...$, fourth zero of $J_0(4\beta/\omega)$; and \textbf{(f)} $4\beta/\omega=27.49...$, ninth zero of $J_0(4\beta/\omega)$. Note the  changes of scales. Shown here are the formula of Eq.~(\ref{Psimple}) (solid blue line) and the numerical solution (dashed black line), these two being completely indistinguishable.
Bottom line: far off-resonance $\omega=100\omega_0$ expectation value of $\sigma_x$ for a system starting in the $|\!\up\rangle$ state at $t=0$ with: \textbf{(g)} $4\beta/\omega=2.404...$, first zero of $J_0(4\beta/\omega)$; \textbf{(h)} $4\beta/\omega= 11.79...$, fourth zero of $J_0(4\beta/\omega)$; and \textbf{(i)} $4\beta/\omega=27.49...$, ninth zero of $J_0(4\beta/\omega)$. Note the  changes of scales in $\langle \sigma_x\rangle$. Shown here are the formula of Eq.~(\ref{SimpleSigmax}) (solid blue line) and the numerical solution (dashed black line), these two being indistinguishable.
%As seen in \textbf{(a)} and \textbf{(b)}, 
%Eq.~(\ref{ReturnProb}) goes well beyond the Bessel function expansion performed %in standard CDT analysis (which applies to situations such as \textbf{(c)}), %capturing the dynamics nearly flawlessly over the entire range of values for $%\omega/\omega_0$.
}
\label{fig:returnprob}
\vspace{-3mm}
\end{center}
\end{figure*}

Concretely, we get a closed form expression for the evolution operator $\mathsf{U}(t)$ at the 0th order of the accelerated Neumann expansion of the path-sum solution that leads to perfect or near-perfect fits for any physical quantity of interest both on and off CDT resonances. See Appendix~\ref{AccNeumann} for details of the calculations. For example, the return probability to the $|\!\up\rangle$ state is found to be
\begin{align}
&P^{(acc,0)}_{\up\to\up}(t)=\left|\cos \left(\frac{2 \beta}{\omega}  \sin (\omega t)\right)+e^{-\frac{1}{2} i t \omega _0}-1\right.\label{ReturnProb}\\
   &\left.+\int_0^t i \omega _0 e^{-\frac{1}{2} i \tau  \omega _0} \sin ^2\left(\frac{\beta}{\omega}  \big(\sin (\omega \tau)-\sin (\omega t)\big)\right) \, d\tau\right|^2,\nonumber
\end{align}
This formula becomes exact when either $\omega_0\to0$ or $\beta\to 0$, as expected from the acceleration procedure. In general, it provides excellent approximations when $\beta/\omega_0$ is large, see Fig.~(\ref{fig:returnprob} a, b, c).   
%\begin{align}
%P^{(acc,0)}_{\up\to\down}&=\left|-i \sin \left(\frac{2 \beta}{\omega}  \sin %(\omega t)\right)\right.\label{PUltraStrong}\\
%&\hspace{-2mm}\left.-\frac{\omega_0}{2} \int_0^t e^{\frac{i}{2}\omega_0 \tau} \sin
%   \left(\frac{2 \beta}{\omega}  \big(\sin (\omega\tau )-\sin (\omega t)\big)\right)d\tau   \right|^2.\nonumber
%\end{align}
The remaining integral in $P^{(acc,0)}_{\up\to\up}(t)$ has no closed form but can be evaluated explicitely via an infinite series of sines and cosines with Bessel coefficients (Appendix \ref{AccNeumann}). This expansion also indicates that the time-average of the return probability is 
\begin{equation}\label{TimeAverage}
\langle P^{(acc,0)}_{\up\to\up}(t)\rangle_t=\frac{1}{2}\left(1+J_0\left(\frac{4 \beta}{\omega}\right)\right),
\end{equation}
which is exactly $1/2$ on CDT resonances where $J_0(4\beta/\omega)=0$, consistent with the current understanding of CDT. To be more precise let us study CDT directly by considering the states $|\psi_\pm\rangle=\frac{1}{\sqrt{2}}(|\!\up\rangle\pm|\!\down\rangle)$. The probability of transition between these states, denoted $P_{\psi_-\to\psi_+}\!(t)$, is found in the situation where $\omega_0\ll (\beta/\omega)^{1/2}$, as (Appendix~\ref{AccNeumann})
\begin{align}
 &P^{(acc,0)}_{\psi_-\to\psi_+}(t)=\frac{\omega_0^2}{4}\left(\int_0^t\sin\Big(\frac{4\beta}{\omega}\big(\sin(\omega t)-\sin(\omega \tau)\big)\Big) d\tau \right)^2\nonumber \\
 &\hspace{5mm}+\frac{\omega_0^2}{4}\left(\int_0^t\cos\Big(\frac{4\beta}{\omega}\big(\sin(\omega t)-\sin(\omega \tau)\big)\Big) d\tau \right)^2.\label{Psimple}
\end{align}
This expression \emph{flawlessly} reproduces the numerical solution in its finest details, details which had hitherto not been captured with such accuracy \cite{Lu2012}. Minimizing the time-average of this formula confirms that the CDT condition is exactly $J_0(4\beta/\omega)=0$, i.e. this is not changed by the non-perturbative corrections. Mathematically, the reason for this is simple: the $J_0$ function is quadratically dominant over the other terms of the Bessel-series expansion of Eq.~(\ref{Psimple}) because it stems from the sole term of that expansion which does not depend on $\tau$ in both integrals. %This is because the $J_0$ term is quadratically dominant over the other terms in the Bessel series expansion of the integrals appearing in Eq.~(\ref{Psimple}). 
%On such CDT resonances the time-average of $  
%Furthermore, Eq.~(\ref{Psimple}) stems from a more involved formula for $P_{\psi_-\to\psi_+}(t)$ that remains valid when $\omega_0$ is not very small compared to $(\beta/\omega)^{1/2}$, see  Appendix~\ref{AccNeumann}.

While these results are as expected from the standard theory of CDT, it not so for all physical quantities. Consider for example, the expectation value of $\sigma_x$ for a system initially prepared in the $|\!\up\rangle$ state. As observed by \cite{Hanggi2000}, $\langle \sigma_x\rangle$ presents anomalous fluctuations on CDT resonances, a fact that was interpreted as a hallmark of and resulting from a crossing Floquet states. This interpretation is in fact not correct. Indeed, at order 0 of the accelerated expansion of the path-sum solution we get (Appendix~\ref{AccNeumann}),
%\begin{align*}
%&\langle \sigma_x\rangle^{(acc,0)}=\omega_0\int_0^t\cos \left(\frac{1}{2}\omega_0 \tau%\right) \sin \left(\frac{4 \beta}{\omega}  \sin
%   (\omega \tau )\right)d\tau\nonumber\\&\hspace{2mm}+2\omega_0 \sin
 %  \left(\frac{1}{4} \omega _0 t\right)\times\nonumber\\&\hspace{4mm}\int_0^t \sin
  % \left(\frac{1}{4} \omega _0 (t-2 \tau )\right)
   %\sin \left(\frac{4 \beta}{\omega }  (\sin (\omega t)-\sin
   %(\omega \tau   ))\right)d\tau.
   %\end{align*}
   %Considering that 
   when $\omega_0 \ll (\beta/\omega)^{1/2}$,% this simplifies to 
   \begin{equation}
   \langle \sigma_x\rangle^{(acc,0)}= \omega_0\int_0^t \sin \left(\frac{4 \beta}{\omega}  \sin
   (\omega \tau )\right)d\tau.\label{SimpleSigmax}
\end{equation}
This simple expression fits once again absolutely flawlessly with the numerically computed expectation $\langle \sigma_x\rangle$, see Fig.~(\ref{fig:returnprob} d, e, f). Now evaluating the integral remaining in Eq.~(\ref{SimpleSigmax}) via Bessel functions shows that the time average of $\langle \sigma_x\rangle$ is
\begin{equation*}
\langle\langle \sigma_x\rangle^{(acc,0)}\rangle_t=\frac{2 \omega_0}{\omega}\sum_{n=0}^\infty J_{2n+1}\left(\frac{4\beta}{\omega}\right)\frac{1}{2n+1},
\end{equation*}
whose extrema are reached whenever 
\begin{equation}\label{Struve}
1-\frac{\pi}{2}\,\pmb{H}_1\left(\frac{4\beta}{\omega}\right)=0,
\end{equation}
with $\pmb{H}_1(.)$ the first Struve function. Remarquably, the difference $\Delta_n$ between the location of the $n$th zero of  $J_0(.)$ and of the $n$th zero of Eq.~(\ref{Struve}) tends asymptotically to 0 as $\Delta_n\sim 1/(2\pi n)$ for $n\gg 1$. This asymptotics develops quite quickly: while $\Delta_1\simeq 0.4$,  already $\Delta_2\simeq 0.03$. The fact that the anomalous fluctuations in the expectation value of $\sigma_x$ peak at the zeroes of Eq.~(\ref{Struve}) rather than on CDT resonances is confirmed by the numerical simulations. This analysis indicates that while $\langle \sigma_x\rangle$ does indeed seem to fluctuate the most on CDT resonances, it is in fact not true and the phenomenon driving these fluctuations is subtly different from that behind CDT.\\[-.5em]

These results demonstrate the power of various expansions of the path-sum solution, enabling very precise and hitherto unequaled analytical analysis of subtle phenomena, e.g. $P_{\psi_-\to\psi_+}(t)$ is on the order of $10^{-5}$ on CDT resonances and is fitted to within machine precision by the formulae provided. This is not because of special features of the Bloch-Siegert Hamiltonian. Rather, the path-sum approach is generally valid for any  driving field, as showed by the general solution provided in  \S\ref{Gen2}. This same solution is valid for dissipative non-Hermitian operators \cite{Sergi2013}, and will always be amenable to analytic Neumann and accelerated Neumann expansions, should it lack a closed form. 

\section{Few- to many-body Hamiltonians}
\subsection{Few-body, $N\!\!>\!\!2\,$-level Hamiltonians}
The path-sum approach is by no mean limited to two-level systems: e.g. solutions to all time-dependent $3\times3$ and $4\times 4$ Hamiltonians are readily available and will be presented in a future work.  %As an example, the construction of the finite path-sum continued fraction is detailed in Fig.~\ref{fig:PSstructure} for a complete $3\times 3$ matrix. Note how 
The number of steps in the exact solution is always finite and the terms involved get progressively simpler because of the "descending ladder principle" (see Fig.~\ref{fig:PSstructure} e).
%Generally, as long as the system is finite, any entry of the evolution operator is given by a finite number of $\ast$-inverses and we are only limited here by the growth of the path-sum continued fraction, itself controlled by the structure of the Hamiltonian. 

For many body systems $N\gg1$, a further problem appears, namely the exponential growth in the size of the Hamiltonian. While path-sum does not, in itself, solve the challenges posed by this well-known scaling, it offers tools to manage it via its scale invariance properties, which we now briefly present as we will use it to treat a many-body molecular system from NMR.%, to which we now turn.

\subsection{Scale invariance}
Path-sums stem from formal resummations of families of walks. This principle does not depend on what those walks represent. In particular, it remains unchanged by the nature of the evolving system. To exploit this observation, consider a more general type of system histories made of temporal successions of orthogonal vector spaces $\tilde{h}:~V_1\mapsto V_2\mapsto V_3 \cdots$. 
Physically such histories can describe an evolving subsystem, such as a group of protons in a large molecule.
Mathematically they correspond to walks on a coarse-grained representation of the quantum state space, a subgraph $\tilde{\G}_t$ of $\G_t$. To see this,  take a complete family of orthogonal spaces, i.e. $\bigoplus_{i=1} V_i=V$, where $V$ is the entire quantum state space. To each $V_i$ associate a vertex $v_i$ and give the edge $v_i\mapsto v_j$ the time-dependent weight $\mathsf{P}_{V_j}.\mathsf{H}(t)\,.\mathsf{P}_{V_i}$. Here $\mathsf{P}_{V_k}$ is the projector onto $V_k$. Observe then that these edge weights are generally non-Abelian. Yet, because path-sums fundamentally retain the order and time of the transitions in histories when performing resummations of walks, this setup poses no further difficulty. It follows that the submatrix $\mathsf{P}_{V_j}.\mathsf{U}(t',t)\,.\mathsf{P}_{V_i}$ of the evolution operator is again given as a matrix-valued branched continued fraction of finite depth and breadth. While the shape of this fraction depends on the particular choice of vector spaces, its existence and convergence properties do not. If the vector spaces are chosen so that the shape of the fraction itself is unchanged, and such a choice is always possible, then the path-sum formulation is truly invariant under scale changes in the quantum state space.
 
An immediate consequence of scale-invariance is that there is always a path-sum calculation rigorously relating the global evolution of a system to that of any ensemble of its subsystems, such as clusters of spins in a large molecule (see below). In this scheme, we can evolve each subsystem \textit{separately} from one-another using any preferred method (Magnus, Floquet, path-sum, Zassenhaus for short times etc.); only to then combine these isolated evolutions \textit{exactly} via a path-sum to generate the true system evolution. 
\begin{figure*}[!t]
\begin{center}
\includegraphics[width=.85\textwidth]{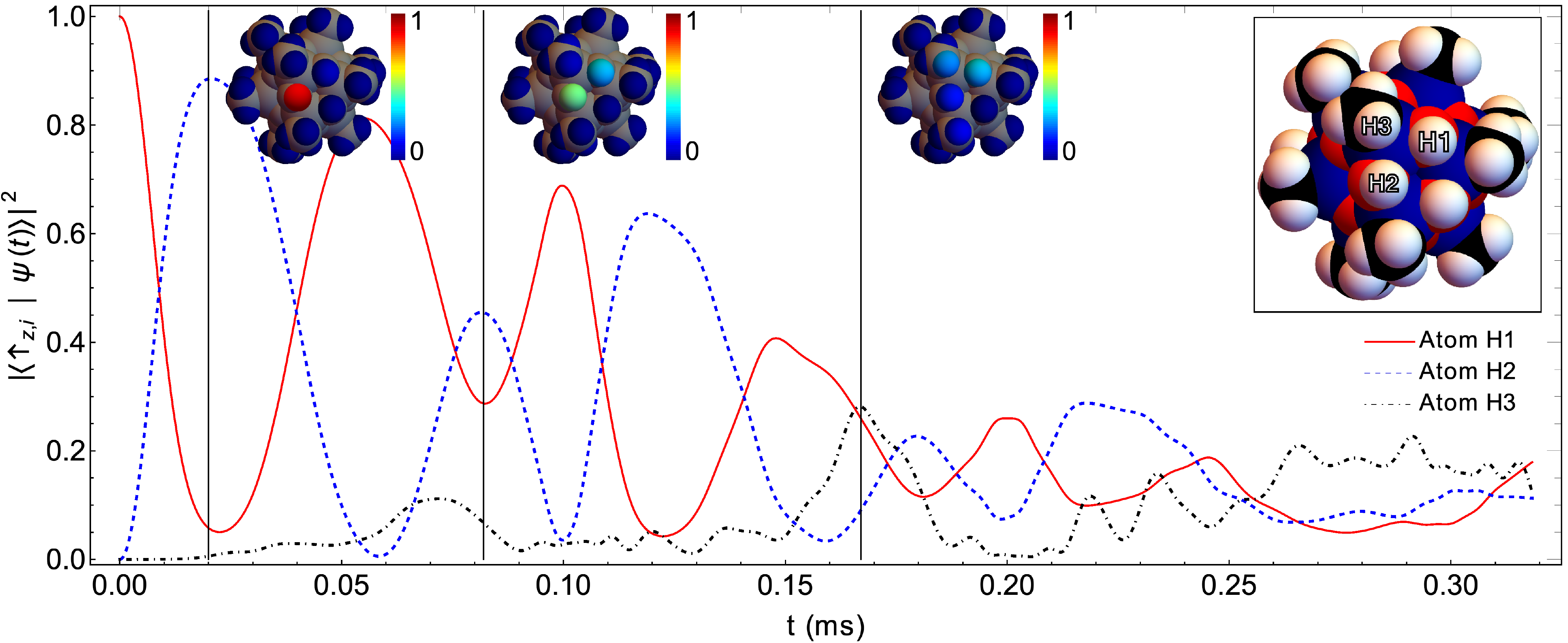}
\caption{\textbf{Analytical spin-diffusion} on a cationic tin oxo-cluster with $N=42$ protons (shown in inset) submitted to the time-dependent high-field dipolar Hamiltonian under MAS (rotor angular velocity $\omega_r=2\pi\times10$kHz). The figure shows the time evolution of the probability $|\langle\psi(t)|\up_{z,i}\rangle|^2$ of finding a spin-up along $z$ on proton $i$ for three protons: a hydroxyl proton H1 (solid red line), on which the excitation starts; a nearby hydroxyl proton H2 (dashed blue line); and a methyl proton H3 (dot-dashed black line).}
\label{fig:MeSnPlot}
\end{center}
\end{figure*}  
While thorough exploitation of the scale-invariance property is beyond the scope of this work, we demonstrate below how it can be used to tackle many-body Hamiltonians, with an emphasis on examples from NMR, i.e. 42 spins coupled by the homonuculear dipolar interaction and spin diffusion under MAS.
\section{Large molecule in NMR}
We now turn to the general problem of determining the temporal dynamics of spin diffusion  as effected by the time-dependent high-field dipolar Hamiltonian for $N$ homonuclear spins:
\begin{equation}\label{HII}
\mathsf{H}^{II} = \sum_{i,j}\frac{1}{2}\omega_{ij}(t)\big(3I_{iz}I_{jz}-\mathsf{I}_{i}\,.\,\mathsf{I}_{j}\big),%=\sum_{i,j}\mathsf{H}_{ij}^{II}.
\end{equation}
where the interaction amplitude $\omega_{ij}(t)$ is time-dependent due to the MAS rotation, see \ref{IntApp} for more details. 
We consider a cationic tin oxo-cluster $\big[(\text{MeSn})_{12}\text{O}_{14}(\text{OH})_6\big]^{2+}$ \cite{Reyes2002} exhibiting $N = 42$ protons belonging to hydroxyl and methyl groups, see Fig.~\ref{fig:MeSnPlot}. This structure is idealised and exhibits the main characteristics of already synthesised clusters (distances, angles, crystal packing). 
%All atomic coordinates as well as selected internuclear distances are given in SI. 
%We consider the full homonuclear dipolar coupling Hamiltonian as well as the chemical shift Hamiltonian.  tin oxo-cluster, 
The methyl groups are supposed fixed as is the case at low temperature, although this is no requirement of the path-sum method and methyl rotations can be tackled. A single orientation of the molecule towards the principal magnetic field $B_0$ is considered.
%, while the extension to a powder could be easily obtained by using averaging procedures over the crystallites \cite{eden1998} or an expression derived from a Fokker-Planck equation \cite{edwards2013}. 
Path-sum yields analytical expressions for the entries of the evolution operator because the computational complexity of the calculations can be made to be only linear in the system size $N$ depending on the initial state. We stress that this is due primarily to the peculiar structure of the high-field dipolar Hamiltonian, which allows for a particularly efficient usage of the scale-invariance and graphical nature of path-sums. In particular, we do not claim to have solved the general many-body problem: there will be Hamiltonians for which this procedure cannot circumvent the exponential explosion of the state space. The methodology we employed is presented below, after the results.
\begin{figure*}[!t]
\includegraphics[width=1\textwidth]{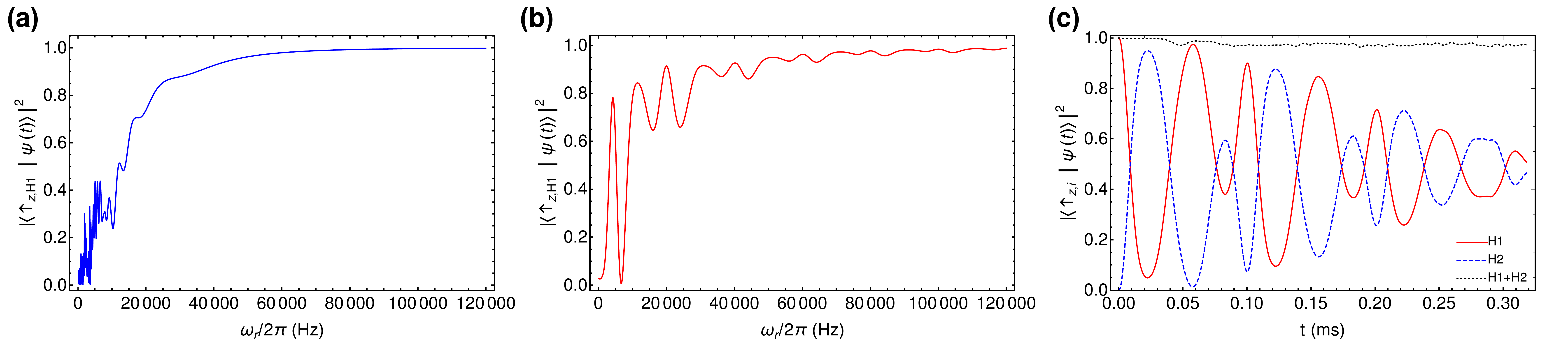}
\begin{center}
\caption{\textbf{Return probability} for the spin excitation on the initial hydroxyl proton H1 as a function of $\omega_R$ (one plot point every 20Hz): \textbf{(a)} after a fixed time of $t=0.05$ms, a situation exhibiting numerous peaks for small $\omega_R$ values that are not all resolved on this picture; and \textbf{(b)} after two rotor periods $t=2\times (2\pi/\omega_R)$. \textbf{(c)} Probability of finding the spin excitation on hydroxyl protons H1 (solid red line) or H2 (dashed blue line) as a function of time for $\omega_R=10$kHz and with a very strong offset of roughly 30 ppm at 1.5 GHz on all protons except H1 and H2. The total probability of being either on H1 or H2 (dotted black line) never goes below $\simeq0.94$ over 3 rotor periods.}
\label{fig:OmegaROffset}
\end{center}
\vspace{-5mm}
\end{figure*}

In Fig.~\ref{fig:MeSnPlot} and Movie~2 (See Supplemental Material at [URL will be inserted by
publisher] for this movie), $\omega_r$ is fixed at $2\pi\times10~$kHz and the initial up-spin is located on a hydroxyl proton, denoted H1. During the first $0.15$ms time period (or 1.5 rotor period), an oscillation is observed between two close hydroxyl protons H1 and H2, followed by a partial transfer to the closest methyl group ($t \gtrsim 0.15$ms), in particular proton H3. Inside the methyl entity, the frequency of exchange is much faster as the three protons are subjected to much stronger dipolar couplings.
In Fig.~\ref{fig:OmegaROffset}(a,b) and Movies~1, 3, 4, 5 and 6 for $\omega_r/2\pi = 5,\,20,\, 40,\, 60$ and $120$ kHz ((See Supplemental Material at [URL will be inserted by
publisher] for these movies), the return-probability to spin 1 is expressed as a function of $\omega_r$ and can be described analytically. These results provide an exact justification to recently proposed approximations in the context of the $^1$H line dependence under ultra-fast MAS \cite{Sternberg2018}.    
Finally in Fig.~\ref{fig:OmegaROffset}(c), strong offsets (roughly 30 ppm at 1.5 GHz, currently the highest magnetic field available for high resolution solid state NMR purposes) %\cite{Gan2017, Bonhomme2018} 
were added to all protons H$_i$, except the two hydroxyl protons H$_{1,2}$ (see inset of Fig.~\ref{fig:MeSnPlot} for identification). As the chemical shift offset corresponds simply to $I_{z,i}$ operators, the solution of the spin diffusion problem remains analytical by using path-sum. For strong offsets, spin diffusion is quenched. All of these results are in perfect agreement with experimental observations related to spin diffusion in NMR.

\subsection{Setting up the path-sum: methodology} 
\subsubsection{State-space reduction techniques}
Simulating many-body quantum systems on classical computers is doomed to be an impossible task, barring the use of approximations. A general class of such approximations, called state-space reduction techniques, bypass the exponential computational hurdle by considering only the most relevant corners of the quantum state-space that the system is likely to explore. But path-sum is, first and foremost, a mathematical technique for analytically solving systems of coupled linear differential equations with non-constant coefficients. This holds regardless of what this system means and how it came about. \emph{Therefore, path-sum can be used in conjunction with all state-space reduction techniques}, as these intervene earlier in selecting the system to be considered. 

In the present work, which focuses on path-sum, we achieve the desired reduction by choosing the initial density matrix $\rho(0)$ to be a pure state with a small number $k$ of up- or down-spins.  Indeed, since the high-field Hamiltonian of Eq.~(\ref{HII}) conserves this number at all times, the discrete graph structure $\mathcal{G}_t$ encoding the quantum state space for path-sum consists of exactly $N$ disconnected components, of sizes $\binom{N}{k}\sim N^k$ when $k\ll N$. Hence, the computational cost of finding the evolution operator using a path-sum here is $O\big(N^k)$, i.e. \textit{linear} in $N$ for a single initial up-spin. % as %\cite{Butler2009, Brusch1997}. 
This procedure is different from approximate state space truncations approaches \cite{Brusch1997,Butler2009,Dumez2010}, since here the Hamiltonian rigorously enforces the state-space partition. As a result, our calculations retain quantum correlations of up to $N$ spins. More general initial density matrices $\rho(0)$ may be approximated with polynomial cost on expanding them over pure states with $k\ll N$. In the sector of the quantum space with a single up-spin, the difficulty is thus solely due to the time-dependent nature of the Hamiltonian. The evolution operator is then strictly analytical for static experiments and analytically soluble using path-sums for MAS experiments.  Physically, the time-dependent high-field dipolar Hamiltonian of Eq.~(\ref{HII}) implements a continuous time quantum random walk of the spin on the molecule. This interpretation remains true in the presence of more than one initial up-spin, with the caveat that further interactions happen when quantum walkers meet. 

%Preliminary observations related to spin chains (see SI) suggest that the results obtained by starting from a pure state initial density matrix give a very good first approximation to more general initial mixed states, as usually encountered in standard room temperature NMR. This particular point will be studied in-depth in a separate work.

\subsubsection{Dynamics at the molecular scale}
As stated above, the sector of the quantum state space that needs to be considered for an initial pure state with a single up-spin is of dimension $N$. 
This reduces the problem of calculating the evolution operator to (analytically) solving an $N\times N$ system of coupled linear differential equations with non-constant coefficients. Since, in principle, all pairs of spins interact directly, this system is full. Consequently, if no further partition of the Hamiltonian is used, the graph $\mathcal{G}_t$ on which path-sum is to be implemented is the complete graph on $N$ vertices, which entails a huge (yet finite) number of terms in the path-sum continued fraction. The vast majority of these give negligible contributions to the overall dynamics however, because of the scales of the interactions involved: one may therefore build up the path-sum continued fraction by brute force, progressively including longer cycles until convergence of the solution is obtained. 

An alternative, physically motivated approach appealing once more to scale-invariance nonetheless appears preferable as it yields further insights in the temporal dynamics.  
%In order to control this consistently, we removed simple cycles from the path-sum continued fraction whose contribution was below some cutoff $1/\Lambda$ of the maximum cycle, and studied convergence as $\Lambda\to\infty$. See Fig.~\ref{} for an illustration. 
% 
%We also implemented an alternative approach further using path-sum's scale invariance. Indeed
First, remark that at least one further non-trivial partition of the Hamiltonian is quite natural in the case of the cationic tin oxo-cluster: that which puts together all spins belonging to the same methyl or 3 hydroxyls groups. Mathematically, this is equivalent to seeing the Hamiltonian as a $14\times 14$ matrix with matrix valued entries, each of size $3\times 3$. Then there is a path-sum continued fraction expressing any $3\times 3$ block of the global evolution operator $\mathsf{U}(t',t)$ in terms of the "small" Hamiltonians of the corresponding proton groups.\\[-1em] 
\begin{figure*}[!t]
\begin{center}
\includegraphics[width=\textwidth]{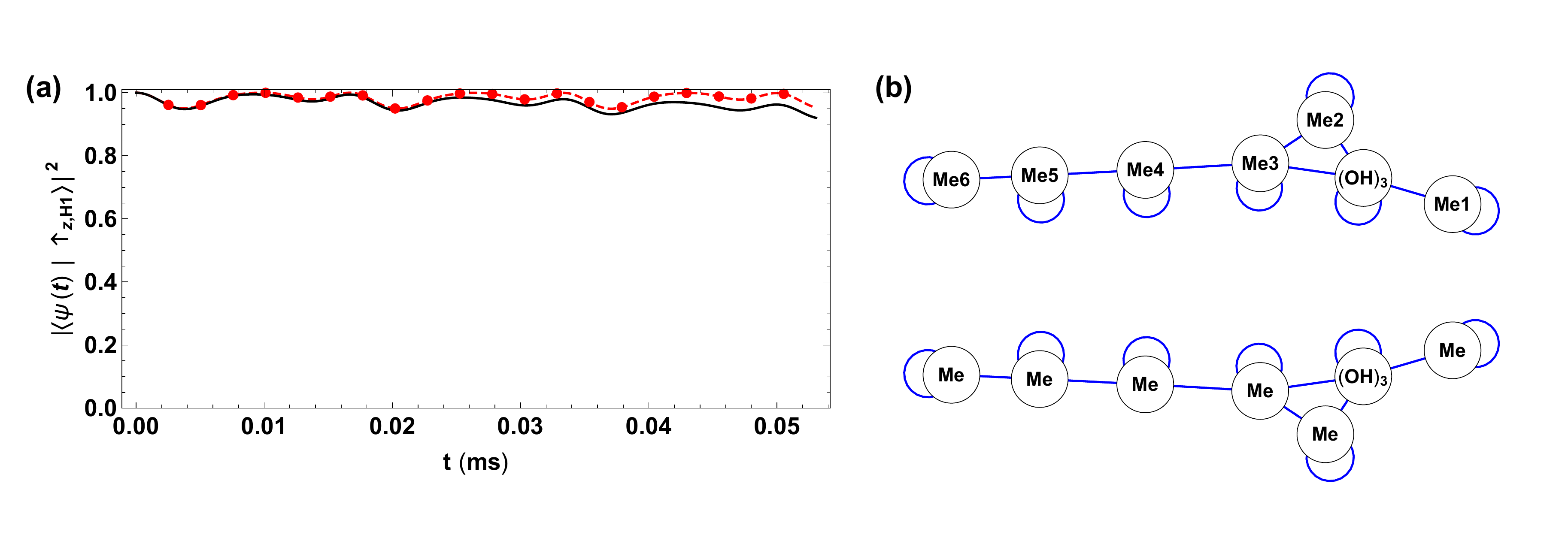}
\vspace{-12mm}
\caption{\textbf{Building the path-sum on the cationic tin oxo-cluster.} \textbf{(a)} Probability of return of the spin excitation on a hydroxyl proton H1 shown on Fig.~\ref{fig:MeSnPlot} as a function of time for $\omega_r=2\pi\times 60$ kHz over 3 rotor periods: (i) solution with no cut-off (solid black line, identical with $\Lambda>100$), (ii) analytical approximation with low interaction cut-off $\Lambda=40$ (dashed red line) and (iii) further approximation obtained upon setting $\Sigma_5$ to zero (red points). \textbf{(b)} Discrete structure $\tilde{\mathcal{G}}_t$ of the quantum state space as seen by path-sum when $\Lambda=40$ and corresponding to the equations given in the text for $\mathsf{U}_{\text{(OH)}_3}$. 
Edges and self-loops correspond to inter-and intra-group interactions, respectively. The adjacency matrix of this graph is the $14\times14$ Hamiltonian with $3\times 3$ matrix valued entries evoked in the text. 
Thus, the shape of the graph is essentially that of the molecule at the methyl and group of 3 hydroxyls level. 
It comprises two disconnected pathways for spin diffusion corresponding to the opposite sides of the cationic tin oxo-cluster which become connected for higher cut-off values $\Lambda>42$.
}
\label{fig:MBconvergence}
\end{center}
\end{figure*}

At this point the path-sum continued fraction is already quite manageable without further approximations, but we can gain additional (analytical) insights into the spin dynamics by removing inter-group interactions weaker than a chosen cut-off value $I_{B,B'}/\Lambda$, with $I_{B,B'}$ the maximum inter-group interaction. Here $B$ indices mean "block". The value of $\Lambda$ is itself controlled by convergence of the overall solution. This procedure sends some off-diagonals blocks of the Hamiltonian to 0, giving $\tilde{\mathcal{G}}_t$ a non-trivial topology which reveals the molecular structure at the methyl and 3 hydroxyls scale, as experienced by the spin excitation during diffusion. See Fig.~\ref{fig:MBconvergence} for an illustrative example, with $\omega_r=2\pi\times 60$ kHz and $\Lambda=40$. The corresponding path-sum continued fraction takes on the topology of the molecule and establishes mathematically the main pathways taken by the spin excitation:
%\frac{1}{\mathsf{Id}_\ast-\mathsf{H}_{\text{Me}}-\mathsf{H}_{\text{MM}}\ast\Sigma_{1}\ast\mathsf{H}_{\text{MM}}},\\
\begin{align*}
&\mathsf{U}_{\text{(OH)}_3}=\int_t^{t'}\!\!\Big(\mathsf{Id}_\ast+i\mathsf{H}_{\text{(OH)}_3}+\mathsf{H}_{\text{(OH)}_3\text{Me}_1}\ast\Gamma_{1}\ast\mathsf{H}_{\text{Me}_1\text{(OH)}_3}\\
&\hspace{-1.5mm}+\mathsf{H}_{\text{(OH)}_3\text{Me}_2}\ast\Sigma_{2}\ast\mathsf{H}_{\text{Me}_2\text{(OH)}_3}+\mathsf{H}_{\text{(OH)}_3\text{Me}_3}\ast\Sigma_{3}\ast\mathsf{H}_{\text{Me}_3\text{(OH)}_3}\\
&\hspace{2mm}-i\,\mathsf{H}_{\text{(OH)}_3\text{Me}_2}\ast\Gamma_2\ast\mathsf{H}_{\text{Me}_2\text{Me}_3}\ast\Sigma_3\ast\mathsf{H}_{\text{Me}_3\text{(OH)}_3}\\
&\hspace{2mm}-i\,\mathsf{H}_{\text{(OH)}_3\text{Me}_3}\ast\Sigma_3\ast\mathsf{H}_{\text{Me}_3\text{Me}_2}\ast\Sigma_2\ast\mathsf{H}_{\text{Me}_2\text{(OH)}_3}\Big)^{\ast-1}\!\!\!(\tau,t)d\tau,
\end{align*}
where e.g. $\mathsf{H}_{\text{(OH)}_3\text{Me}_3}\ast\Sigma_3\ast\mathsf{H}_{\text{Me}_3\text{Me}_2}\ast\Sigma_2\ast\mathsf{H}_{\text{Me}_2\text{(OH)}_3}$ is the weight of the triangle $\text{(OH)}_3\mapsto \text{Me}_2\mapsto \text{Me}_3 \mapsto \text{(OH)}_3$ on $\tilde{\mathcal{G}}_t$ (Fig.~\ref{fig:MBconvergence} b). In these expressions, $\mathsf{Id}_\ast=1_\ast \mathsf{Id}_{3\times 3}$, the $\Sigma_j$ are given by
\begin{align*}
&\Sigma_2 = \frac{1}{\mathsf{Id}_\ast+i\mathsf{H}_{\text{Me}_2}+\mathsf{H}_{\text{Me}_{2}\text{Me}_{3}}\ast\Sigma_3\ast\mathsf{H}_{\text{Me}_{3}\text{Me}_{2}}},\\
&\Sigma_3=\frac{1}{\mathsf{Id}_\ast+i\mathsf{H}_{\text{Me}_3}+\mathsf{H}_{\text{Me}_3\text{Me}_4}\ast\Sigma_4\ast\mathsf{H}_{\text{Me}_4\text{Me}_3}},\\
&\Sigma_4=\frac{1}{\mathsf{Id}_\ast+i\mathsf{H}_{\text{Me}_4}+\mathsf{H}_{\text{Me}_4\text{Me}_5}\ast\Sigma_5\ast\mathsf{H}_{\text{Me}_5\text{Me}_4}},\\
&\Sigma_5 = \frac{1}{\mathsf{Id}_\ast+i\mathsf{H}_{\text{Me}_5}+\mathsf{H}_{\text{Me}_5\text{Me}_6}\ast\Gamma_6\ast\mathsf{H}_{\text{Me}_6\text{Me}_5}},
\end{align*}
and $\Gamma_j$ designates the isolated evolution of the $j$th methyl group, i.e. 
\begin{align*}
&\Gamma_j = \frac{1}{\mathsf{Id}_\ast+i\mathsf{H}_{\text{Me}_j}}.
%&\Sigma_1= \frac{1}{\mathsf{Id}_\ast+i\mathsf{H}_{\text{Me}_1}},\\
%&= \frac{1}{\mathsf{Id}_\ast+i\mathsf{H}_{\text{Me}_3}}
\end{align*}
These results illustrate again the ``descending ladder principle'' evoked in Fig.~\ref{fig:PSstructure}. Here, all inverses are $\ast$-inverses and $\mathsf{U}_{\text{(OH)}_3}$ is the $3\times 3$ block of the global evolution operator giving the probability amplitudes over a group of 3 hydroxyls. $\mathsf{H}_{\text{Me}_x}$ and $\mathsf{H}_{\text{(OH)}_3}$ are the Hamiltonians of isolated methyl and of a group of 3 hydroxyls, respectively. Similarly, $\mathsf{H}_{\text{Me}_i\text{Me}_j}$ is the interaction between neighbouring methyls and $\mathsf{H}_{\text{Me}_i\text{(OH)}_3}$ the interaction between a methyl and a group of 3 hydroxyls. 

The reader may notice that the shape taken by the continued fraction for $\mathsf{U}_{\text{(OH)}_3}$ is immediately related to that of the graph $\tilde{\mathcal{G}}_t$ (Fig.~\ref{fig:MBconvergence}(b)), with each term of the fraction being the weight of a fundamental cycle of the graph. This close, transparent, association between the mathematical form of the solution and the physical problem allows for physically motivated and better controlled approximations. For example, setting $\Sigma_5$ to zero so that $\Sigma_4\equiv \Gamma_4$ in the above solution is immediately understood to mean that one removes the possibility for the spin to diffuse to the remote methyl groups Me5 and Me6 before coming back to the initial group of 3 hydroxyls, an excellent approximation (see Fig.~\ref{fig:MBconvergence}(a), red points to be compared to the red dashed line).

%This yields any $3\times 3$ block of the global evolution operator in the sector of the quantum state space with a single up-spin.  
Finally, we remark that our choice of partition is not mathematically necessary. For example, larger blocks may be employed equally well or one may form blocks with protons scattered throughout the molecule. In principle, path-sum's scale-invariance guarantees that any choice, if properly implemented, leads to the same solution. In practice however there is a trade-off between the size of the manipulated blocks and the complexity of the path-sum continued fraction. We do not know in general how to choose the best partition according to this trade-off but it seems that physically motivated partitions are a good starting point.
%, although it  seems to us that non-trivial partition better reveals the spin exchanges at the molecular scale. 

\section{Conclusion}
In this contribution, we have demonstrated an entirely novel approach to the problem of finding compact and exact expressions for the evolution operators of quantum dynamical systems driven by time-varying Hamiltonians. 
%In fact, the potential of application of path-sum exceeds Hermitian matrices since it can be applied to any finite matrix\footnote{Technically, we need the matrix entries to be bounded functions of time over the time-interval of interest in order to guarantee super-exponential convergence of the Neumann series for the $\ast$-inverses \cite{Giscard2015}} and even extends to infinite ones under additional assumptions such as translation invariance. 
As illustrated in Figure~\ref{fig:PSstructure}, path-sum calculations always involve a ``descending ladder'' of progressively simpler quantities yielding  the exact solution after a finite number of steps. This is in strong contrast with traditional perturbation techniques (Magnus expansion, Floquet theory) which, when carying out analytically, invariably lead to infinite series and an ``ascending ladder'' of increasingly intricate quantities, such as Magnus series' nested commutators.
Most importantly, the solutions provided by path-sums are always analytically accessible, e.g. through Neumann expansions. 
%Relaxation and decoherence processes will be taken explicitly into account by extending path-sum to the Liouvillian space.%Mathematically, since these solutions can involve special functions, no better form for them can in general be expected. Path-sum may alternatively be implemented fully numerically by exploiting the properties of the separable linear Volterra equations of the second kind, which the method produces. This option has not been used in this contribution.

As a fundamental and illustrative example, we used path-sum to solve the Bloch-Siegert problem---related to the action of the counter-rotating component of the radio-frequency field---at any order. %We anticipate that further complex, and currently unsolved problems, involving $3\times 3$, $4\times 4, \cdots,$ matrices will be solved using path-sums as well. As a remarquable example, the exact evolution of the macroscopic magnetisation $M$ in a strong magnetic field $B_0$ under the action of any $B_1(t)$ radio-frequency field (a $3\times3$ problem) will be presented in a separate contribution. 
We analytically studied the spin diffusion effected by the homonuclear dipolar coupling Hamiltonian of NMR acting on a large molecule, starting from a pure state initial density matrix. In general, on many-body systems, we are facing two kinds of "explosive" computational problems: (i) one, quantum in nature, related to the exponential size of the quantum state space; and (ii) one, graph theoretical in nature, related to the time required to construct the path-sum continued fraction, in particular if $\mathcal{G}_t$ is large and not sparse. Issue (ii) can be managed with partitions and path-sum's scale invariance and is further  tackled with the implementation of a Lanczos path-sum algorithm \cite{GiscardPozza2019}. This algorithm naturally exploits matrix sparsity, benefits from path-sum's ``descending ladder'' principle and was designed with a numerical outlook. It can, in principle, get excellent approximations after only a few iterations, equivalent to truncating a path-sum continued fraction in sufficient depth to reach the desired accuracy. This algorithm is best understood as an extension to time-ordered exponentials of modern numerical procedures for the computation of ordinary matrix exponentials.
%Lanczos path-sum relies on an exact transformation of the Hamiltonian into a time-dependent tri-diagonal matrix. 
The first issue (i) is fundamental to quantum mechanics and its management inherently depends on the problem at hand. Here, path-sum has the advantage that it works in conjunction with any state-space reduction technique. For the homonuclear dipolar coupling Hamiltonian, we bypassed the problem upon choosing certain initial pure states. The scale invariance of path-sum offers further flexibility, as it allows to separately evolve chosen subsystems only to then combine all such evolutions in a globally exact way.

These results call for a discussion on the nature of the solutions sought after by physicists and mathematicians alike. A general assumption seems to be that an acceptable/interesting analytical solution to a problem has been found if and only if it can be presented with a finite number of symbols and pre-existing  functions. We think this is a restrictive if missleading expectation. For example, a Bessel or a Heun function solution would be considered `satifactory' when both are actually algebraically transcendant, known and understood from the equations they solve and from explicit series expansions involving simpler objects. It seems that at least in some cases our perception of mathematical objects may be biased by facts as simple as their having a name, yet the sine integral function $\text{Si}(x)=\int \sin(x)/x dx$ is no more undisputedly  analytical than $\int \exp(\sin(x)/x)) dx$. 
We think that one cannot and should not ask a general purpose analytical method for solving systems of coupled linear differential equations with variable coefficients any more than what there is to be found: i) finding, in a finite number of steps, an \emph{explicit} differential or integral equation involving only one unknown function to be determined; and ii) providing an unconditionally convergent mean of expanding the solution as a series of some kind, be it Taylor, Neumann, accelerated Neumann or other. We may add the requirement that, iii) all calculations  should be feasible \emph{analytically}, i.e. without giving numerical values to all parameters involved. Should one of these criterion fail to be met, a purely numerical strategy would surely be more interesting. But if all of these demands are indeed satisfied, we may analyse the situation in greater depth and details than possible with numerical computations. This is exemplified by the CDT analysis provided here, the analytical formula for $\langle\sigma_x\rangle$ revealing slight deviations from the expected zeroes of the Bessel $J_0$ function.

With these understandings in place, we think that path-sum opens
%In the case of translation-invariant spin-chains and more general Bethe lattices driven by generic long-range Hamiltonians, 
%we show in SI that this allows path-sum to provide \textit{fully exact} expression for the evolution operator, even in the limit $N\to\infty$, although exploiting these solutions remains a formidable challenge. 
an entire new field of research is now open for the NMR and wider physics communities.\\

%\matmethods{Please describe your materials and methods here. This can be more than one paragraph, and may contain subsections and equations as required. Authors should include a statement in the methods section describing how readers will be able to access the data in the paper. 
%
%\subsection*{Subsection for Method}
%Example text for subsection.
%}
%
%\showmatmethods{} % Display the Materials and Methods section

\begin{acknowledgments}
C. Bonhomme thanks Dr. F. Ribot for communicating the molecular data pertaining to the cationic tin oxo-cluster. P.-L. Giscard is supported by the Agence Nationale de la Recherche grant ANR-19-CE40-0006. P.-L. Giscard is also grateful for the financial support from the Royal Commission for the Exhibition of 1851 over the period 2015--2018, during which time the present research was started.
\end{acknowledgments}

\appendix

\section{Remarks on the state-of-the-art}\label{AppA}
While reviewing the state-of-the-art in the course of the present work, it appeared to us that a very vast corpus of research had accumulated on quantum dynamics driven by time-varying Hamiltonians. A host of special solutions have been found and numerous betterments of existing techniques have been developed. Some of these are recent enough that we could not cover them in our introduction, such as the flow equation approach to periodic Hamiltonians \cite{Vogl2019}. It seems that a proper review article on the subject is urgently needed to gather all results and remedy the pitfalls of our modest introduction. 

Following publication of the preprint of the present article, it was suggested to us that path-sum may be related to Haydock's recursion method for calculating electronic states \cite{haydock1991}. While both approaches share the same outlook of recursively resumming Feynman diagrams via path-resummations, Haydock's method relies on fundamentally commutative mathematics, in particular determinants, which do not extend to the general setting required by ordered exponentials and scale invariance. Instead, it is possible that lifting Haydock's approach to non-commutativity using Gelfand's quasi-determinants \cite{Gelfand1997} would lead to path-sum.

\section{Bloch Siegert dynamics}\label{2LevelExplicit}
In this appendix, we detail the calculation process for the transition probability $P_{\up\to\down}$ at order 3 of the Neumann expansion of the exact path-sum solution. We work on resonance $\omega_0=\omega$ as this yields more compact expressions and also because this situation is the most challenging mathematically. Indeed, precisely when $\omega=\omega_0$ the kernel $K_\up$ has terms that are linear in time and which slow down convergence of $ P^{(n)}_{\up\to\down}(t)$ to $P_{\up\to\down}(t)$ (see \S\ref{BSPS}).

As explained in the main text, at order 3 we have $P^{(3)}_{\up\to\down}(t)=|U^{(3)}(t)_{\down \up}|^2$ with
$$
U^{(3)}(t)_{\down\up}=-2i\beta\int_{0}^{t}\int_{0}^{\tau_1} \cos(\omega\tau_1)e^{i\omega_0 \tau_1}G^{(3)}_\up(\tau_0,0)d\tau_0d\tau_1,
$$
see Eq.~(\ref{Udu}) of the main text. Here $G^{(3)}_\up$ is the third order Neumann expansion of the path-sum solution, i.e.
\begin{widetext}
\begin{align*}
G^{(3)}_\up(t,0)=&\delta(t)+K_\up(t,0)+\int_0^t K_\up(t,\tau_1)K_\up(\tau_1,0)d\tau_1+\int_0^t\int_{\tau_1}^t K_\up(t,\tau_2)K_\up(\tau_2,\tau_1)K_\up(\tau_2,0)d\tau_2d\tau_1,\\
=& \delta(t)-\frac{\beta^2}{\omega }e^{-i\omega t}\cos (\omega t ) \left( -i e^{2 i \omega t }+2 \omega t+i\right) \\
&+\frac{\beta^4}{24 \omega ^3} e^{-3 i \omega t }\cos (
   \omega t ) \Big(3 i e^{6 i  \omega t }+6 e^{4 i \omega t } (-2i \omega^2 t^2  +2  \omega  t+i)+e^{2 i \omega t } (  8\omega^3 t^3 +12i\omega^2 t^2+12\omega t -15 i)-12  \omega t+6 i\Big)\\
   &+\frac{\beta ^6 }{960 \omega ^5}e^{-5 i  \omega t }\cos (\omega t) \Big(e^{10i\omega t}5i+e^{8i\omega t}(-60 i \omega ^2 t^2+150 \omega t +150 i)+e^{6i\omega t}(40 i \omega ^4 t^4-80 \omega ^3 t^3+420 \omega t +150 i)\\
   &\hspace{35mm}+e^{4i \omega t}(-16 \omega ^5 t^5-40 i \omega ^4 t^4-160 \omega ^3 t^3+180 i \omega ^2 t^2+360 \omega t -380 i)\\
   &\hspace{35mm}+e^{2i\omega t}(80 \omega^3 t^3-120 i  \omega^2 t^2-180  \omega t +45 i)-30 \omega t +30 i\Big). 
\end{align*}
\end{widetext}
%Here we recall that the on-resonance kernel is
%$$
%K_\up(t',t)=\frac{\beta^2}{\omega}  \left(i e^{2 i \omega t' }-i e^{2 i \omega t }-2 \omega  (t'-t)\right)e^{-i \omega t' } \cos (\omega t' ).
%$$

Taken together, these calculations give the transition probability at the third Neumann order as 
$$
P^{(3)}_{\up\to\down}(t)=\sum_{k=0}^{7}\sin(2k\omega t) S^{(3)}_{2k}(\beta,t)+\cos(2k\omega t) C^{(3)}_{2k}(\beta,t),
$$
in accordance with Eq.~(\ref{Pseries}) of the main text.
Here we have
\begin{widetext}
\begin{scriptsize}
\begin{align*}
S^{(3)}_0(\beta,t)=&0,\\
S^{(3)}_2(\beta,t)=&\frac{\beta ^{14} t^{13}}{3628800 \omega }+\frac{\beta ^{14} t^{11}}{53760 \omega ^3}-\frac{\beta ^{12} t^{11}}{50400 \omega }-\frac{\beta ^{14} t^9}{129024 \omega ^5}-\frac{209 \beta ^{12} t^9}{241920 \omega ^3}+\frac{41 \beta ^{10} t^9}{60480 \omega }-\frac{229 \beta ^{14} t^7}{645120
   \omega ^7}+\frac{11 \beta ^{12} t^7}{3840 \omega ^5}+\frac{17 \beta ^{10} t^7}{960 \omega ^3}-\frac{4 \beta ^8 t^7}{315 \omega }\\
   &-\frac{4219 \beta ^{14} t^5}{6635520 \omega ^9}-\frac{125 \beta ^{12} t^5}{9216 \omega ^7}-\frac{13 \beta ^{10} t^5}{192 \omega ^5}-\frac{\beta ^8 t^5}{6 \omega
   ^3}+\frac{2 \beta ^6 t^5}{15 \omega }+\frac{1615 \beta ^{14} t^3}{73728 \omega ^{11}}+\frac{22873 \beta ^{12} t^3}{331776 \omega ^9}+\frac{265 \beta ^{10} t^3}{1152 \omega ^7}+\frac{7 \beta ^8 t^3}{16 \omega ^5}+\frac{2 \beta ^6 t^3}{3 \omega ^3}-\frac{2 \beta ^4 t^3}{3 \omega
   }\\
   &-\frac{1249411 \beta ^{14} t}{21233664 \omega ^{13}}-\frac{17357 \beta ^{12} t}{110592 \omega ^{11}}-\frac{9419 \beta ^{10} t}{27648 \omega ^9}-\frac{211 \beta ^8 t}{384 \omega ^7}-\frac{21 \beta ^6 t}{32 \omega ^5}-\frac{3 \beta ^4 t}{4 \omega ^3}+\frac{\beta ^2 t}{\omega },\\
S^{(3)}_4(\beta,t)=&-\frac{19 \beta ^{14} t^{11}}{4838400 \omega ^3}-\frac{73 \beta ^{14} t^9}{1935360 \omega ^5}+\frac{97 \beta ^{12} t^9}{483840 \omega ^3}+\frac{799 \beta ^{14} t^7}{1161216 \omega ^7}+\frac{47 \beta ^{12} t^7}{80640 \omega ^5}-\frac{61 \beta ^{10} t^7}{13440 \omega ^3}-\frac{27401 \beta
   ^{14} t^5}{6635520 \omega ^9}-\frac{629 \beta ^{12} t^5}{46080 \omega ^7}+\frac{\beta ^{10} t^5}{384 \omega ^5}+\frac{\beta ^8 t^5}{20 \omega ^3}\\
   &+\frac{82439 \beta ^{14} t^3}{3538944 \omega ^{11}}+\frac{18683 \beta ^{12} t^3}{331776 \omega ^9}+\frac{1061 \beta ^{10} t^3}{13824 \omega
   ^7}-\frac{\beta ^8 t^3}{12 \omega ^5}-\frac{\beta ^6 t^3}{4 \omega ^3}-\frac{1125889 \beta ^{14} t}{21233664 \omega ^{13}}-\frac{213157 \beta ^{12} t}{1769472 \omega ^{11}}-\frac{8675 \beta ^{10} t}{55296 \omega ^9}+\frac{7 \beta ^8 t}{288 \omega ^7}+\frac{7 \beta ^6 t}{32 \omega
   ^5}+\frac{3 \beta ^4 t}{8 \omega ^3},\\
S^{(3)}_6(\beta,t)=&\frac{29 \beta ^{14} t^9}{1935360 \omega ^5}-\frac{493 \beta ^{14} t^7}{1658880 \omega ^7}-\frac{121 \beta ^{12} t^7}{241920 \omega ^5}+\frac{5039 \beta ^{14} t^5}{4423680 \omega ^9}+\frac{385 \beta ^{12} t^5}{55296 \omega ^7}+\frac{19 \beta ^{10} t^5}{2880 \omega ^5}+\frac{4643 \beta ^{14}
   t^3}{1769472 \omega ^{11}}-\frac{5065 \beta ^{12} t^3}{221184 \omega ^9}-\frac{27 \beta ^{10} t^3}{512 \omega ^7}-\frac{5 \beta ^8 t^3}{144 \omega ^5}\\
   &-\frac{480511 \beta ^{14} t}{31850496 \omega ^{13}}+\frac{6845 \beta ^{12} t}{884736 \omega ^{11}}+\frac{1513 \beta ^{10} t}{18432 \omega
   ^9}+\frac{229 \beta ^8 t}{2304 \omega ^7}+\frac{5 \beta ^6 t}{96 \omega ^5},\\
S^{(3)}_8(\beta,t)=&-\frac{911 \beta ^{14} t^7}{46448640 \omega ^7}+\frac{6877 \beta ^{14} t^5}{13271040 \omega ^9}+\frac{47 \beta ^{12} t^5}{122880 \omega ^7}-\frac{1187 \beta ^{14} t^3}{442368 \omega ^{11}}-\frac{1721 \beta ^{12} t^3}{331776 \omega ^9}-\frac{133 \beta ^{10} t^3}{55296 \omega ^7}+\frac{14657
   \beta ^{14} t}{5308416 \omega ^{13}}+\frac{2153 \beta ^{12} t}{221184 \omega ^{11}}+\frac{1237 \beta ^{10} t}{110592 \omega ^9}+\frac{35 \beta ^8 t}{9216 \omega ^7},\\
S^{(3)}_{10}(\beta,t)=&\frac{41 \beta ^{14} t^5}{4423680 \omega ^9}-\frac{103 \beta ^{14} t^3}{589824 \omega ^{11}}-\frac{19 \beta ^{12} t^3}{221184 \omega ^9}+\frac{1339 \beta ^{14} t}{3538944 \omega ^{13}}+\frac{53 \beta ^{12} t}{98304 \omega ^{11}}+\frac{\beta ^{10} t}{6144 \omega ^9},\\
S^{(3)}_{12}(\beta,t)=&-\frac{5 \beta ^{14} t^3}{3538944 \omega ^{11}}+\frac{703 \beta ^{14} t}{63700992 \omega ^{13}}+\frac{7 \beta ^{12} t}{1769472 \omega ^{11}},\\
S^{(3)}_{14}(\beta,t)=&\frac{\beta ^{14} t}{21233664 \omega ^{13}}.
\end{align*}
\end{scriptsize}
\end{widetext}
Now on to the $C^{(3)}_{2k}$ functions:
\begin{widetext}
\begin{scriptsize}
\begin{align*}
C^{(3)}_0(\beta,t)=&\frac{5809339 \beta ^{14}}{509607936 \omega ^{14}}+\frac{1327 \beta ^{12}}{73728 \omega ^{12}}+\frac{25787 \beta ^{10}}{442368 \omega ^{10}}+\frac{10315 \beta ^8}{55296 \omega ^8}+\frac{157 \beta ^6}{384 \omega ^6}+\frac{7 \beta ^4}{16 \omega ^4}+\frac{\beta ^2}{2 \omega ^2}+\frac{\beta
   ^{14} t^{14}}{25401600}+\frac{\beta ^{14} t^{12}}{290304 \omega ^2}-\frac{\beta ^{12} t^{12}}{302400}+\frac{11 \beta ^{14} t^{10}}{358400 \omega ^4}\\
   &-\frac{59 \beta ^{12} t^{10}}{302400 \omega ^2}+\frac{41 \beta ^{10} t^{10}}{302400}
   -\frac{1007 \beta ^{14} t^8}{1548288 \omega
   ^6}-\frac{137 \beta ^{12} t^8}{322560 \omega ^4}+\frac{23 \beta ^{10} t^8}{4480 \omega ^2}-\frac{\beta ^8 t^8}{315}+\frac{48229 \beta ^{14} t^6}{7962624 \omega ^8}+\frac{1441 \beta ^{12} t^6}{138240 \omega ^6}-\frac{31 \beta ^{10} t^6}{3840 \omega ^4}-\frac{\beta ^8 t^6}{15 \omega
   ^2}+\frac{2 \beta ^6 t^6}{45}\\
   &-\frac{302327 \beta ^{14} t^4}{10616832 \omega ^{10}}-\frac{93581 \beta ^{12} t^4}{1327104 \omega ^8}-\frac{55 \beta ^{10} t^4}{4608 \omega ^6}+\frac{7 \beta ^8 t^4}{48 \omega ^4}+\frac{5 \beta ^6 t^4}{12 \omega ^2}-\frac{\beta ^4 t^4}{3}+\frac{660727 \beta
   ^{14} t^2}{10616832 \omega ^{12}}+\frac{41143 \beta ^{12} t^2}{294912 \omega ^{10}}+\frac{21131 \beta ^{10} t^2}{110592 \omega ^8}-\frac{133 \beta ^8 t^2}{384 \omega ^6}\\
   &-\frac{\beta ^6 t^2}{2 \omega ^4}-\frac{\beta ^4 t^2}{\omega ^2}+\beta ^2 t^2,\\
   C^{(3)}_2(\beta,t)=&\frac{22781 \beta ^{14}}{7077888 \omega ^{14}}+\frac{6671 \beta ^{12}}{196608 \omega ^{12}}+\frac{727 \beta ^{10}}{27648 \omega ^{10}}-\frac{541 \beta ^8}{55296 \omega ^8}-\frac{139 \beta ^6}{384 \omega ^6}-\frac{\beta ^4}{2 \omega ^4}-\frac{\beta ^2}{2 \omega ^2}-\frac{\beta ^{14}
   t^{12}}{1036800 \omega ^2}-\frac{\beta ^{14} t^{10}}{120960 \omega ^4}+\frac{\beta ^{12} t^{10}}{17280 \omega ^2}-\frac{421 \beta ^{14} t^8}{1935360 \omega ^6}\\
   &+\frac{\beta ^{12} t^8}{4480 \omega ^4}-\frac{\beta ^{10} t^8}{640 \omega ^2}+\frac{221941 \beta ^{14} t^6}{39813120 \omega
   ^8}+\frac{383 \beta ^{12} t^6}{138240 \omega ^6}-\frac{\beta ^{10} t^6}{320 \omega ^4}+\frac{\beta ^8 t^6}{45 \omega ^2}-\frac{95713 \beta ^{14} t^4}{2654208 \omega ^{10}}-\frac{82045 \beta ^{12} t^4}{1327104 \omega ^8}+\frac{43 \beta ^{10} t^4}{3072 \omega ^6}-\frac{\beta ^6 t^4}{6
   \omega ^2}+\frac{111757 \beta ^{14} t^2}{1179648 \omega ^{12}}\\
   &+\frac{22853 \beta ^{12} t^2}{147456 \omega ^{10}}+\frac{19945 \beta ^{10} t^2}{110592 \omega ^8}-\frac{7 \beta ^8 t^2}{64 \omega ^6}+\frac{\beta ^6 t^2}{8 \omega ^4}+\frac{\beta ^4 t^2}{2 \omega ^2},\\
   C_4^{(3)}(\beta,t)=&-\frac{1146415 \beta ^{14}}{254803968 \omega ^{14}}-\frac{27865 \beta ^{12}}{589824 \omega ^{12}}-\frac{2849 \beta ^{10}}{27648 \omega ^{10}}-\frac{1439 \beta ^8}{6912 \omega ^8}-\frac{29 \beta ^6}{384 \omega ^6}+\frac{\beta ^4}{16 \omega ^4}-\frac{53 \beta ^{14} t^{10}}{3225600 \omega
   ^4}-\frac{37 \beta ^{14} t^8}{368640 \omega ^6}+\frac{5 \beta ^{12} t^8}{7168 \omega ^4}+\frac{4567 \beta ^{14} t^6}{2488320 \omega ^8}+\frac{47 \beta ^{12} t^6}{69120 \omega ^6}\\
   &-\frac{49 \beta ^{10} t^6}{3840 \omega ^4}-\frac{15503 \beta ^{14} t^4}{1327104 \omega ^{10}}-\frac{3773 \beta
   ^{12} t^4}{165888 \omega ^8}+\frac{73 \beta ^{10} t^4}{4608 \omega ^6}+\frac{5 \beta ^8 t^4}{48 \omega ^4}+\frac{128387 \beta ^{14} t^2}{3538944 \omega ^{12}}+\frac{2639 \beta ^{12} t^2}{36864 \omega ^{10}}+\frac{37 \beta ^{10} t^2}{432 \omega ^8}-\frac{31 \beta ^8 t^2}{192 \omega
   ^6}-\frac{5 \beta ^6 t^2}{16 \omega ^4},\\
   C^{(3)}_6(\beta,t)=&-\frac{521561 \beta ^{14}}{47775744 \omega ^{14}}-\frac{92075 \beta ^{12}}{10616832 \omega ^{12}}+\frac{2915 \beta ^{10}}{221184 \omega ^{10}}+\frac{1565 \beta ^8}{55296 \omega ^8}+\frac{11 \beta ^6}{384 \omega ^6}+\frac{199 \beta ^{14} t^8}{1935360 \omega ^6}-\frac{33029 \beta ^{14}
   t^6}{39813120 \omega ^8}-\frac{373 \beta ^{12} t^6}{138240 \omega ^6}+\frac{3919 \beta ^{14} t^4}{5308416 \omega ^{10}}+\frac{22013 \beta ^{12} t^4}{1327104 \omega ^8}\\
   &+\frac{239 \beta ^{10} t^4}{9216 \omega ^6}+\frac{165683 \beta ^{14} t^2}{21233664 \omega ^{12}}-\frac{1741 \beta ^{12}
   t^2}{73728 \omega ^{10}}-\frac{9593 \beta ^{10} t^2}{110592 \omega ^8}-\frac{\beta ^8 t^2}{12 \omega ^6},\\
   C_8^{(3)}(\beta,t)=&\frac{315173 \beta ^{14}}{509607936 \omega ^{14}}+\frac{407 \beta ^{12}}{110592 \omega ^{12}}+\frac{2261 \beta ^{10}}{442368 \omega ^{10}}+\frac{173 \beta ^8}{55296 \omega ^8}-\frac{5681 \beta ^{14} t^6}{39813120 \omega ^8}+\frac{5029 \beta ^{14} t^4}{3538944 \omega ^{10}}+\frac{2645 \beta
   ^{12} t^4}{1327104 \omega ^8}-\frac{12631 \beta ^{14} t^2}{3538944 \omega ^{12}}-\frac{8077 \beta ^{12} t^2}{884736 \omega ^{10}}-\frac{851 \beta ^{10} t^2}{110592 \omega ^8},\\
   C_{10}^{(3)}(\beta,t)=&\frac{10979 \beta ^{14}}{63700992 \omega ^{14}}+\frac{1067 \beta ^{12}}{3538944 \omega ^{12}}+\frac{37 \beta ^{10}}{221184 \omega ^{10}}+\frac{307 \beta ^{14} t^4}{5308416 \omega ^{10}}-\frac{2351 \beta ^{14} t^2}{7077888 \omega ^{12}}-\frac{145 \beta ^{12} t^2}{442368 \omega ^{10}},\\
   C_{12}^{(3)}(\beta,t)=&\frac{5309 \beta ^{14}}{764411904 \omega ^{14}}+\frac{25 \beta ^{12}}{5308416 \omega ^{12}}-\frac{65 \beta ^{14} t^2}{10616832 \omega ^{12}},\\
   C_{14}^{(3)}(\beta,t)=&\frac{\beta ^{14}}{15925248 \omega ^{14}}.
\end{align*}
\end{scriptsize}
\end{widetext}

%Some sub-series of terms in the $S_{2k}$ and $C_{2k}$ functions clearly yield closed form expressions when taken together, e.g. one can recognise the beginning of expansions of tangents and sine of cosines in the above. However, the transcendent nature of $G_\up$ guarantees that even with such resummations at least some of $S_{2k}$ and $C_{2k}$ will still involve infinite series functions to be summed over. 

All calculations were performed analytically on \textsc{Mathematica}. The notebook generating these results, as well as any desired higher order of the Neumann expansion of the exact path-sum solution is available for download at \url{http://www-lmpa.univ-littoral.fr/~plgiscard/}. Everytime the order is increased by one, e.g. from $P^{(3)}_{\up\to\down}(t)$ to $P^{(4)}_{\up\to\down}(t)$, each expression above gains new high order terms while four new functions also appear, e.g. $S^{(4)}_{16}(\beta,t)$, $S^{(4)}_{18}(\beta,t)$, $C^{(4)}_{16}(\beta,t)$ and $C^{(4)}_{18}(\beta,t)$ all enter $P^{(4)}_{\up\to\down}(t)$).

\section{Accelerated Neumann series}\label{AccNeumann}
Suppose that we are given a function or matrix of two times $K(t',t)=K_1(t',t)+K_2(t',t)$ such that in some sense $K_1$ is much larger than $K_2$. Suppose further that we are interested in the solution of the linear Volterra integral equation of the second kind $G(t',t)=\delta+K\ast G= \delta(t'-t)+\int_t^{t'} K(t',\tau)G(\tau,t)d\tau$, as will always be the case when expanding the exact path-sum solution to quantum dynamical problems at any scale.  

Instead of expanding $G$ as usual, 
$
G(t',t)=\delta+\sum_n K^{\ast n}
$,
one can exploit the fact that $K_1$ is dominant over $K_2$ to accelerate convergence of the Neumann expansion by expressing $G$ in terms of the solutions $G_{i}$ of the ``individual'' Volterra equations $G_i= \delta+K_i\ast G_i$. More specifically one gets
$$
G=\left(\sum_n T^{\ast n}\right)\ast G_1\ast G_2= G_1\ast G_2+ T\ast G_1\ast G_2 + \cdots,
$$
where $T=\delta(t'-t)-G_1\ast G_2+G_1\ast G_2\ast (K_1+K_2)$, see \cite{VolterraGiscard} for details. Since $T^{\ast0}=\delta(t'-t)$, the 0th order term of the accelerated expansion is then simply the $\ast$-product of the solutions of the individual Volterra equations:
$$
G^{(acc,0)}(t',t)=T^{\ast 0}\ast G_1\ast G_2 = \int_t^{t'} G_1(t',\tau)G_2(\tau,t)d\tau.
$$
This is particularly well suited to physical situations where a certain parameter dominates over the others: not only because the so-obtained expression for $G$ is greatly improved, but also because in general the individual $G_i$ are known exactly. Furthermore, this acceleration procedure continues to hold for any number of kernels $K_i$ \cite{VolterraGiscard}.

Taking the Bloch-Siegert Hamiltonian of Eq.~(\ref{HBS1}) as an example, let us use  path-sum's scale invariance to work in the trivial situation where we have single subsystem, namely the entire system itself. Then, we get that 
$$
\mathsf{U}(t)=\int_0^t \mathsf{G}(\tau,0) d\tau,
$$
with $G$ the solution of the matrix-valued linear integral Volterra equation of the second kind with matrix kernel
$\mathsf{K}=\mathsf{K}_1+\mathsf{K}_2$, where
\begin{align*}
\mathsf{K}_1(t)&=-2i\beta\begin{pmatrix}0&\cos(\omega t)\\\cos(\omega t)&0\end{pmatrix},\\ \mathsf{K}_2(t)&=-i\omega_0\begin{pmatrix}1/2&0\\0&-1/2\end{pmatrix}.
\end{align*}
The ultra-strong coupling regime $\beta/\omega_0\gg 1$ thus corresponds to the situation described above as $K_1$ dominates $K_2$. Since furthermore both $\mathsf{G}_i$ are immediately accessible as
$$
\mathsf{G}_i(t',t)=\delta(t'-t)\mathsf{Id}+\mathsf{K}_i(t') \exp\left(\int_{t}^{t'}\mathsf{K}_i(\tau)d\tau\right),
$$
we get $\mathsf{G}^{(acc,0)}$ easily 
%\begin{widetext}
%\begin{align*}
%\mathsf{U}^{(acc,0)}(t)&=\delta(t'-t)\mathsf{Id}+\\
%&\begin{pmatrix}\beta  \left(-2+i \omega_0 t\right) \cos (\omega t ) \sin \left(\frac{2 \beta}{\omega} 
%   \sin (\omega t )\right)-\frac{1}{2} i \omega _0 e^{-\frac{1}{2} i 
%   \omega_0 t} & \beta  \left(\omega_0 t-2 i\right) \cos (\omega t ) \cos
%   \left(\frac{2 \beta}{\omega} \sin (\omega t )\right) \\
% -\beta  \left(\omega_0 t+2 i\right) \cos (\omega t ) \cos \left(\frac{2 \beta}{\omega}  
%   \sin (\omega t )\right) &    
%     \beta \left(-2-i\omega_0 t \right) \cos (\omega t ) \sin
%   \left(\frac{2 \beta}{\omega}   \sin (\omega t )\right)+\frac{1}{2} i\omega _0 e^{\frac{1}{2} i \omega_0 t }
%   \end{pmatrix},
%\end{align*}
%\end{widetext}
and integrating it with respect to $t$ yields
\begin{widetext}
%\begin{small}
\begin{align*}
\mathsf{U}^{(acc,0)}(t)=&\begin{pmatrix}
 \cos \left(\frac{2 \beta}{\omega}  \sin (\omega t)\right)+e^{-\frac{1}{2} i  \omega_0 t}-1 & -i \sin \left(\frac{2 \beta}{\omega}  \sin (\omega t)\right) \\
 -i \sin \left(\frac{2 \beta}{\omega}  \sin (\omega t)\right) & \cos \left(\frac{2 \beta}{\omega}  \sin (\omega t)\right)+e^{\frac{1}{2}i  \omega_0 t}-1 \\
\end{pmatrix}\\
&\hspace{10mm}+\int_0^t \begin{pmatrix}
 i \omega_0 e^{-\frac{1}{2} i   \omega_0 \tau} \sin ^2\left(\frac{2\beta}{\omega}  \big(\sin (\omega \tau)-\sin (\omega t)\big)\right) & -\frac{1}{2}
   \omega_0 e^{\frac{1}{2}i \omega_0 \tau} \sin \left(\frac{4 \beta}{\omega}  \big(\sin (\omega \tau)-\sin (\omega t)\big)\right) \\
 \frac{1}{2} \omega_0 e^{-\frac{1}{2} i   \omega_0 \tau} \sin \left(\frac{4 \beta}{\omega}  \big(\sin (\omega \tau)-\sin (\omega t)\big)\right) & -i
   \omega_0 e^{\frac{1}{2}i \omega_0 \tau} \sin ^2\left(\frac{2\beta}{\omega}  \big(\sin (\omega \tau)-\sin (\omega t)\big)\right) \\
\end{pmatrix} d\tau.
\end{align*}
%\end{small}
\end{widetext}
Higher orders of the accelerated expansion of the path-sum solution are also available although they are not necessary given the machine-precision accuracy with respect to numerical solutions already reached by order 0. The integrals in $\mathsf{U}^{(acc,0)}(t)$ have no closed form but can be determined exactly via standard expansions over Bessel functions since e.g.
\begin{align*}
 &\sin(\alpha+z\sin(\phi))=\\
 &\hspace{5mm}\sin(\alpha)\left(J_0(z)+2\sum_{n=1}^\infty J_{2n}(z) \cos(2n\phi)\right)\\
 &\hspace{7mm}+2\cos(\alpha)\sum_{n=0}^\infty J_{2m+1}(z)\sin((2n+1)\phi).
\end{align*}

The modulus squared of $\mathsf{U}^{(acc,0)}(t)_{11}$ gives Eq.~(\ref{ReturnProb}) of the main text, 
%while e.g. $|\mathsf{U}^{(acc,0)}(t)_{12}|^2$ gives the transition probability  
%\begin{align*}
%P^{(acc,0)}_{\up\to\down}&=\left|-i \sin \left(\frac{2 \beta}{\omega}  \sin (\omega t)\right)\right.\\
%&\hspace{-2mm}\left.-\frac{\omega_0}{2} \int_0^t e^{\frac{i}{2}\omega_0 \tau} \sin
%   \left(\frac{2 \beta}{\omega}  \big(\sin (\omega\tau )-\sin (\omega t)\big)\right)d\tau   \right|^2.
%\end{align*}
while calculating other quantities such as $P_{\psi_-\to\psi_+}(t)$ and $\langle \sigma_x\rangle$ from $\mathsf{U}^{(acc,0)}(t)$ is now a simple task, giving e.g.
\begin{align*}
&\langle \sigma_x\rangle^{(acc,0)}=\omega_0\int_0^t\cos \left(\frac{1}{2}\omega_0 \tau\right) \sin \left(\frac{2 \beta}{\omega}  \sin
   (\omega \tau )\right)d\tau\\&\hspace{2mm}+2\omega_0 \sin
   \left(\frac{1}{4} \omega _0 t\right)\times\\&\hspace{5mm}\int_0^t \sin
   \left(\frac{1}{4} \omega _0 (t-2 \tau )\right)
   \sin \left(\frac{2 \beta}{\omega }  (\sin (\omega t)-\sin
   (\omega \tau   ))\right)d\tau.
\end{align*}
In the regime $\omega_0\ll \omega$, both $\cos(\omega_0 t/2)$ and $\sin(\omega_0 t/4)$ are essentially equal to their initial $t=0$ values, leading to Eq.~(\ref{SimpleSigmax}).

\section{Interaction terms in the high-field dipolar Hamiltonian}\label{IntApp}
We consider the time-dependent high-field dipolar Hamiltonian of Eq.~(\ref{HII}) presented in the main text, with interaction terms under MAS 
$$
\omega_{ij}(t) := \frac{\mu_0 \gamma^2 \hbar}{4\pi r_{ij}^3}\times \frac{1}{2}\xi_{ij}(t),
$$
where $r_{ij}$ is the distance between protons $i$ and $j$ and \cite{slichter1990}
\begin{align*}
\xi_{ij}(t) &:= 2\sqrt{2}\sin(\psi_{ij})\cos(\psi_{ij})\sin(\phi_{ij}+\omega_r t)\\
&\hspace{20mm}+\sin(\psi_{ij})^2 \cos(2\phi_{ij}+2\omega_r t).
\end{align*}
In this expression, $\psi_{ij}$ is the angle between $\vec{ij}$ and the $z$-axis and $\phi_{ij}$ is the angle between $\vec{ij}$ and the $x$-axis for a coordinate system fixed to the sample.
Finally, $\omega_r$ is the angular velocity of the rotor. The raw molecular data pertaining to the cationic tin oxo-cluster is available online on the webpage \url{http://www-lmpa.univ-littoral.fr/~plgiscard/} and included here as a dataset.\\ 

%\bibliography{RMN_PathSum}
%

\end{document}